\documentclass[12pt]{article}
\usepackage[a4paper, total={7in, 10in}]{geometry}
\usepackage[parfill]{parskip}
\usepackage{physics, tensor, float, subcaption}
\usepackage{graphicx}
\graphicspath{ {Plots/} }
\usepackage{jhep-mod}
\usepackage{bm}
\usepackage{soul}
\usepackage{amssymb,amsmath,amsthm}
\usepackage{mathrsfs}
\usepackage[utf8]{inputenc}
\usepackage{enumerate}
\usepackage{bigints}
\usepackage{xcolor}
\usepackage{appendix}
\usepackage{graphicx}
\usepackage{float}
\usepackage{tikz}
\usepackage{setspace}
\usepackage{cancel}
\definecolor{purple}{rgb}{1,0,1}
\definecolor{lime}{HTML}{A6CE39} % needs xcolor
\definecolor{lime}{HTML}{A6CE39}
\newcommand{\orcidicon}{%
	\begin{tikzpicture}
	\draw[lime, fill=lime] (0,0) 
		circle [radius=0.16] 
		node[white] {{\fontfamily{qag}\selectfont \tiny ID}};
	\draw[white, fill=white] (-0.0625,0.095) 
		circle [radius=0.007];
	\end{tikzpicture}
	\hspace{-5mm}
}
\newcommand\orcidThomas{{\href{https://orcid.org/0000-0002-0314-4136}{\orcidicon}}}
\newcommand\orcidAlex{{\href{https://orcid.org/0000-0002-1763-3563}{\orcidicon}}}
\newcommand\orcidMatt{{\href{https://orcid.org/0000-0003-1088-6485}{\orcidicon}}}
\newcommand{\e}{\mathrm{e}}

\begin{document}
%========================================================

%========================================================
%========================================================
\title{\vspace{-25pt}{\huge
Photon spheres, ISCOs, and OSCOs: 
Astrophysical observables for regular black holes with asymptotically Minkowski cores
}}
%========================================================
%========================================================
%========================================================
\author{
\Large
Thomas Berry\!\orcidThomas, Alex Simpson\!\orcidAlex\!, {\sf  and} Matt Visser\!\orcidMatt}
%========================================================
%========================================================
%========================================================
%========================================================
\affiliation{School of Mathematics and Statistics, Victoria University of Wellington, \\
\null\qquad PO Box 600, Wellington 6140, New Zealand.}
%========================================================
%========================================================
\emailAdd{thomas.berry@sms.vuw.ac.nz}
\emailAdd{alex.simpson@sms.vuw.ac.nz}
\emailAdd{matt.visser@sms.vuw.ac.nz}
%========================================================
%========================================================
\abstract{
\vspace{1em}

Classical black holes contain a singularity at their core.
This has prompted various researchers to propose a multitude of modified spacetimes that mimic the physically observable characteristics of classical black holes as best as possible, but that crucially do not contain singularities at their cores.
Due to recent advances in near-horizon astronomy, the ability to observationally distinguish between a classical black hole and a potential black hole mimicker is becoming increasingly feasible.
Herein, we calculate some physically observable quantities for a recently proposed regular black hole with an asymptotically Minkowski core --- the radius of the photon sphere and the extremal stable timelike circular orbit (ESCO). 
The manner in which the photon sphere and ESCO relate to the presence (or absence) of horizons is much more complex than for the Schwarzschild black hole. 
We find situations in which photon spheres can approach arbitrarily close to (near extremal) horizons, situations in which some photon spheres become stable, 
and situations in which the locations of both photon spheres and ESCOs become multi-valued, with both ISCOs (innermost stable circular orbits) and OSCOs (outermost stable circular orbits).  This provides an extremely rich phenomenology of potential astrophysical interest. 

\bigskip
\noindent
{\sc Date:} \\
Monday 31 August 2020; Monday 7 September; \LaTeX-ed \today

\bigskip
\noindent{\sc Keywords}:
regular black hole, Minkowski core, Lambert $W$ function, black hole mimic.

\bigskip
\noindent{\sc PhySH:} 
Gravitation
}

%========================================================
\maketitle
%========================================================
\def\tr{{\mathrm{tr}}}
\def\diag{{\mathrm{diag}}}
\def\cof{{\mathrm{cof}}}
\def\pdet{{\mathrm{pdet}}}
\def\O{{\mathcal{O}}}
\parindent0pt
\parskip7pt
%=====================================================
\section{Introduction}
%=====================================================

Karl Schwarzschild first derived the spacetime metric for the region exterior to a static, spherically symmetric source in 1916~\cite{schwarzschild-original}; only some 50 years later was it properly understood that this spacetime could be extrapolated inwards to describe a black hole. 
Without any loss of generality, any static spherically symmetric spacetime can be described by a metric of the form
\begin{equation}
\dd{s}^2 = - \e^{-2\Phi(r)}\left(1-\frac{2m(r)}{r}\right)\dd{t^2}+\frac{\dd{r^2}}{1-\frac{2m(r)}{r}}+r^2\left(\dd{\theta^2}+\sin^2\theta\dd{\phi^2}\right).
\end{equation}
For the standard Schwarzschild metric one sets \( \Phi(r)=0 \) and \( m(r)=m_0 \).
Over the past century, a vast host of black hole spacetimes, qualitatively distinct from that of Schwarzschild, have been investigated by multiple researchers~\cite{reissner-original, 
weyl-original, nordstrom-original, kerr-original, kerr-newmann, kerr-schild, kerr-intro, 
kerr-book, Baines:2020, Lense-Thirring, vaidya-original1, vaidya-original2, vaidya-original3}.

Furthermore, the field has now grown to not only include classical black holes, but also quantum-modified black holes~\cite{quantum-aspects-of-bhs_book, quantum-bh1, quantum-bh2, quantum-bh3}, regular black holes~\cite{bardeen-rbh, hayward-rbh, frolov-rbh, rbhs-review, rbh-viability}, and various other exotic spherically symmetric spacetimes that are fundamentally different from black holes but mimic many of their observable phenomena (\emph{e.g.} traversable wormholes~\cite{morris-thorne-original, morris-thorne-yurtsever, lorentzian-wormholes, visser-wormhole-examples, visser-surgical, time-machine, baby, visser-kar-dadhich, Kar:2004, poisson-visser, natural, R=0, expmetric, blackbounce, blackbounce2, Lobo:2020}, gravastars~\cite{gravastar-original, gravastars-bh-alternative, visser-gravastars, anisotropic, lobo-gravastars, MartinMoruno:2011, Lobo:2015}, ultracompact objects~\cite{Cunha1,Cunha2}, \textit{etcetera}~\cite{pandora,small-dark-heavy,observability}; see~\cite{visser-phenom-aspects} for an in-depth discussion).
Herein, we investigate a specific model spacetime representing a regular black hole.
That is, a spacetime that has a well-defined horizon structure, but the curvature invariants are everywhere finite.

Investigating black hole mimickers is becoming increasingly relevant due to recent advances in both observational and gravitational wave astronomy. 
Projects such as the Event Horizon Telescope~\cite{eht-1, eht-2, eht-3, eht-4, eht-5, eht-6}, and LIGO~\cite{ligo-detection-papers, grav-wave-observations-wiki}, and the planned LISA~\cite{LISA}, are and will be continuously probing closer to the horizons of compact massive objects (CMOs), and so there is hope that such projects will eventually be able to distinguish between the near-horizon physics of classical black holes and possible astrophysical mimickers~\cite{visser-phenom-aspects}.

The model spacetime investigated in this work is a regular black hole with an asymptotically Minkowski core, as discussed in~\cite{asymptot-mink-core, Berry:2020}.
This is an example of a metric with an exponential mass suppression, and is described by the line element
\begin{equation}
\dd s^2 = -\left(1-\frac{2m\,\e^{-a/r}}{r}\right)\dd{t}^2 + \frac{\dd{r}^2}{1-\frac{2m\,\e^{-a/r}}{r}} + r^2\left(\dd{\theta}^2 + \sin^2\theta \dd{\phi}^2\right).
\label{metric}
\end{equation}
A rather different (extremal) version of this model spacetime, based on nonlinear electrodynamics, has been previously discussed by Culetu~\cite{Culetu:2013}, with follow-up on some
aspects of the non-extremal case in references~\cite{Culetu:2014, Culetu:2015a, Culetu:2015b}. See also~\cite{Junior:2015, Rodrigues:2015}.

Most regular black holes have a core that is asymptotically de~Sitter (with constant positive curvature)~\cite{bardeen-rbh, hayward-rbh, frolov-rbh, rbhs-review}.
However, the regular black hole described by the metric \eqref{metric} has an asymptotically Minkowski core (in the sense that the stress-energy tensor asymptotes to zero).
Such models have some attractive features compared to the more common de~Sitter core regular black holes: the stress-energy tensor vanishes at the core, greatly simplifying the physics in this region; and many messy algebraic expressions are replaced by simpler expressions involving the exponential and Lambert $W$ functions, whilst still allowing for explicit closed form expressions for quantities of physical interest~\cite{asymptot-mink-core}.
Additionally, the results obtained in this work reproduce the standard results for the Schwarzschild metric by letting the parameter \( a \rightarrow 0 \).
Thus, the value of the parameter \( a \) determines the extent of the ``deviation'' from the Schwarzschild spacetime.

\newpage
If $0<a<2m/\e$ then the spacetime described by the metric \eqref{metric} has two horizons located at
\begin{equation}
r_{H^-} = 2m\;\e^{W_{-1}\left(-\frac{a}{2m}\right)}, 
\qquad\hbox{and} \qquad
r_{H^+} = 2m\;\e^{W_{0}\left(-\frac{a}{2m}\right)}.
\label{horizon}
\end{equation}
Here \( W_{-1}(x) \) and   \( W_{0}(x) \) are the real-valued branches of Lambert $W$ function.
We could also write
\begin{equation}
r_{H^-} = {a\over |W_{-1}\left(-\frac{a}{2m}\right)|}, 
\qquad\hbox{and} \qquad
r_{H^+} = {a\over |W_{0}\left(-\frac{a}{2m}\right)|}.
\label{horizon2}
\end{equation}
%\enlargethispage{40pt}

Perturbatively, for small $a$ we have
\begin{equation}
r_{H^+} = 2m -a +\O(a^2),
\label{E:outer-pert}
\end{equation}
nicely reproducing Schwarzschild in the $a\to0$ limit.  For the inner horizon, since $r_{H_-} < 2m$ then
\begin{equation}
r_{H^-} = {a\over \ln(2m/r_{H_-})}
\end{equation}
implies $r_{H^-} < a$, whence we have a strict upper bound given by the simple analytic expression:
\begin{equation}
r_{H^-} < {a\over \ln(2m/a)}.
\label{E:inner-bound}
\end{equation}
Certainly $\lim_{a\to0} r_{H^-}(m,a)=0$ as we would expect to recover Schwarzschild; 
but the form of $r_{H^-}(m,a)$ is not analytic. This bound can also be viewed as the first term in an asymptotic expansion~\cite{Corless:1996} based on (as $x\to 0^+$) 
\begin{equation}
W_{-1}(-x) = \ln(x) + \O(\ln(-\ln(x))) = -\ln(1/x) + \O(\ln(\ln(1/x))). 
\end{equation}
This leads to
\begin{equation}
r_{H^-} = {a\over \ln(2m/a) +\O(\ln(\ln(2m/a)))} 
= {a\over \ln(2m/a)} +
\O\left(a\ln(\ln(2m/a))\over (\ln(2m/a))^2 \right).
\label{E:inner-asymp}
\end{equation}
More specifically (as $a/m\to 0$ or $m/a\to\infty$) 
\begin{equation}
{r_{H^-} \over a} = {1\over \ln(2m/a)} +
\O\left(\ln(\ln(2m/a))\over (\ln(2m/a))^2 \right).
\label{E:inner-asymp2}
\end{equation}

If $a=2m/\e$ then the two horizons merge at $r_H=2m/\e=a$ and one has an extremal black hole. 
If $a >2m/\e$ then there are no horizons, and one is dealing with a regular horizonless extended but compact object, (the energy density peaks at $r=a/4$).

This object could either be extended all the way down to $r=0$, or alternatively be truncated at some finite value of  $r$,  to be used as the exterior geometry for some static and spherically symmetric mass source that \emph{isn't} a black hole. This is potentially useful as a model for planets, stars, \emph{etc.} Consequently, we will also incorporate aspects of the analysis for $a>2m/\e$ as and when required to generate astrophysical observables in the case when equation~(\ref{metric}) is modelling a compact object other than a black hole.

%=====================================================
\section{Geodesics and the effective potential}
%=====================================================

Continuing the analysis of~\cite{asymptot-mink-core}, we will now calculate the location of the photon sphere and extremal stable circular orbit (ESCO) for the regular black hole with line element given by equation \eqref{metric}.
Photon spheres, (or more precisely the closely related black hole silhouettes),  have been recently observed for the massive objects M87 and  Sgr A*~\cite{eht-1,eht-2,eht-3,eht-4,eht-5,eht-6}.  As such they are, along with the closely related ESCOs, practical and useful quantities to calculate for black hole mimickers.

%\clearpage
\enlargethispage{20pt}
We begin by considering the affinely parameterised tangent vector to the worldline of a massive or massless particle in our spacetime \eqref{metric}:
\begin{multline}
g_{\mu\nu}\dv{x^\mu}{\lambda}\dv{x^\nu}{\lambda} = -\left(1-\frac{2m\,\e^{-a/r}}{r}\right)\left(\dv{t}{\lambda}\right)^2 + \left(\frac{1}{1-\frac{2m\,\e^{-a/r}}{r}}\right)\left(\dv{r}{\lambda}\right)^2  \\
+ r^2 \left[ \left(\dv{\theta}{\lambda}\right)^2 + \sin^2\theta \left(\dv{\phi}{\lambda}\right)^2 \right] = \epsilon,
\label{tangent}
\end{multline}
where \( \epsilon \in \{-1,0\} \); with $-1$ corresponding to a massive (timelike) particle and 0 corresponding to a massless (null) particle. (The case $\epsilon=+1$ would correspond to tachyonic particles following spacelike geodesics, a situation of no known physical applicability.) 
Since we are working with a spherically symmetric spacetime, we can set \( \theta = \pi/2 \) without any loss of generality and reduce equation \eqref{tangent} to 
\begin{equation}
g_{\mu\nu}\dv{x^\mu}{\lambda}\dv{x^\nu}{\lambda} = -\left(1-\frac{2m\,\e^{-a/r}}{r}\right)\left(\dv{t}{\lambda}\right)^2 + \left(\frac{1}{1-\frac{2m\,\e^{-a/r}}{r}}\right)\left(\dv{r}{\lambda}\right)^2 + r^2 \left(\dv{\phi}{\lambda}\right)^2 = \epsilon.
\label{tangentreduced}
\end{equation}
Due to the presence of time-translation and angular Killing vectors, we can now define the conserved quantities
\begin{equation}
E = \left(1-\frac{2m\,\e^{-a/r}}{r}\right)\left(\dv{t}{\lambda}\right) \qq{and} L = r^2\left(\dv{\phi}{\lambda}\right),
\end{equation}
corresponding to the energy and angular momentum of the particle, respectively. 
Thus, equation \eqref{tangentreduced} implies
\begin{equation}
E^2 = \left(\dv{r}{\lambda}\right)^2  + \left(1-\frac{2m\,\e^{-a/r}}{r}\right)\left(\frac{L^2}{r^2}-\epsilon\right).
\end{equation}
This defines an ``effective potential'' for geodesic orbits
\begin{equation}
V_\epsilon (r) = \left(1-\frac{2m\,\e^{-a/r}}{r}\right)\left(\frac{L^2}{r^2}-\epsilon\right),
\end{equation}
with the circular orbits corresponding to extrema of this potential.

%========================================================
\section{Photon spheres}
%========================================================

We subdivide the discussion into two topics: First the \emph{existence} of circular photon orbits (photon spheres) and then the \emph{stability} of circular photon orbits.
The discussion is considerably more complex than for the Schwarzschild spacetime, where there is only one circular photon orbit, at $r=3m$, and that circular photon orbit is unstable. Once the extra parameter $a$ is nonzero, and in particular sufficiently large, 
the set of photon orbits exhibits more diversity. 

%========================================================
\subsection{Existence of photon spheres}
%========================================================

For null trajectories we have
\begin{equation}
V_0(r) = \left(1-\frac{2m\,\e^{-a/r}}{r}\right)\frac{L^2}{r^2}.
\end{equation}
So for circular photon orbits
\begin{equation}
V_0'(r_c) = \frac{2L^2}{r_c^5} \left[ m\,\e^{-a/r_c}(3r_c-a)-r_c^2 \right] = 0.
\end{equation}
%\enlargethispage{40pt}

\begin{figure}[!htbp]
\centering
\begin{subfigure}{.5\textwidth}
  \centering
  \includegraphics[width=.95\linewidth]{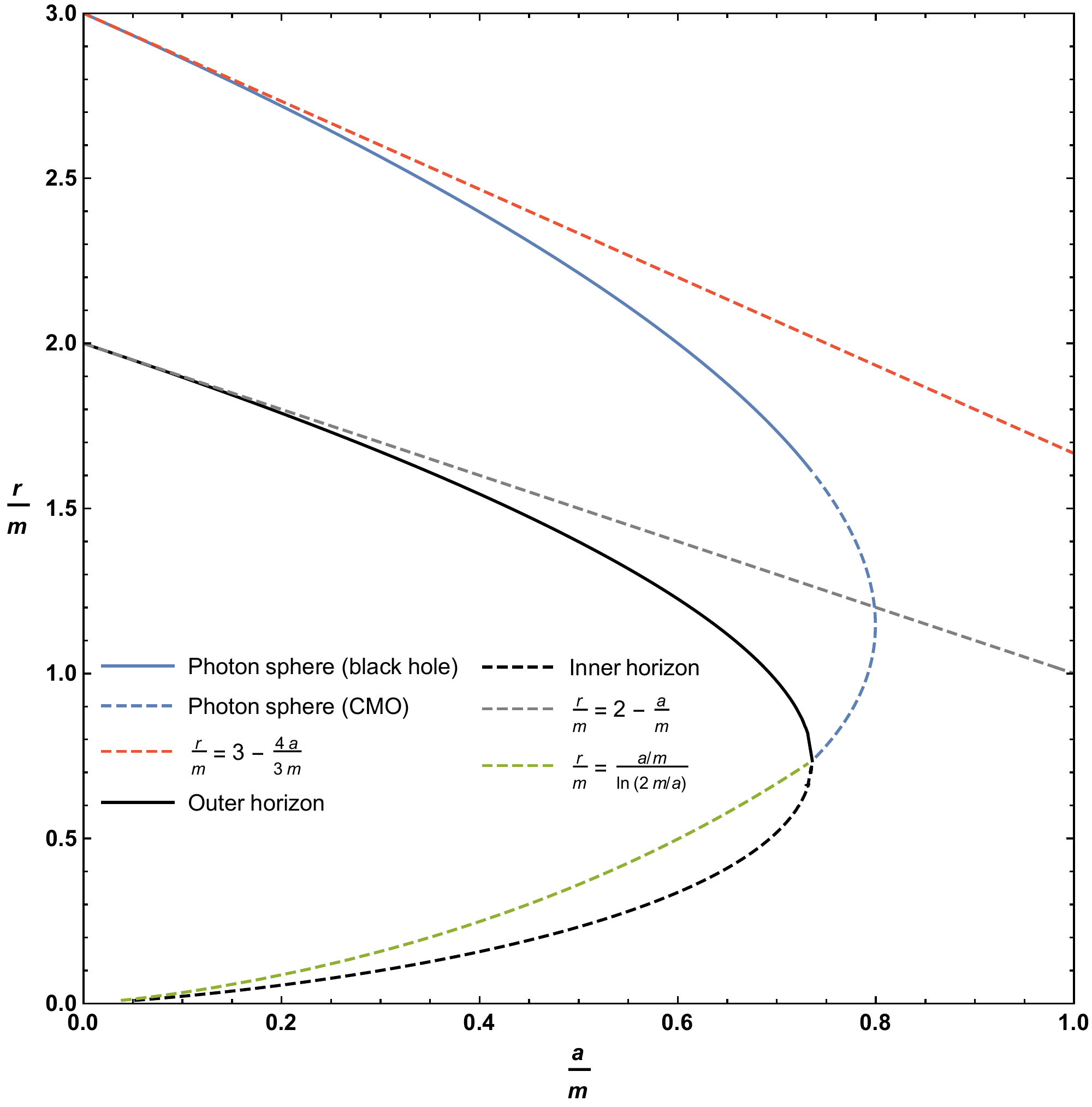}
  \caption{}
  \label{fig:iscom}
\end{subfigure}%
\begin{subfigure}{.5\textwidth}
  \centering
  \includegraphics[width=.95\linewidth]{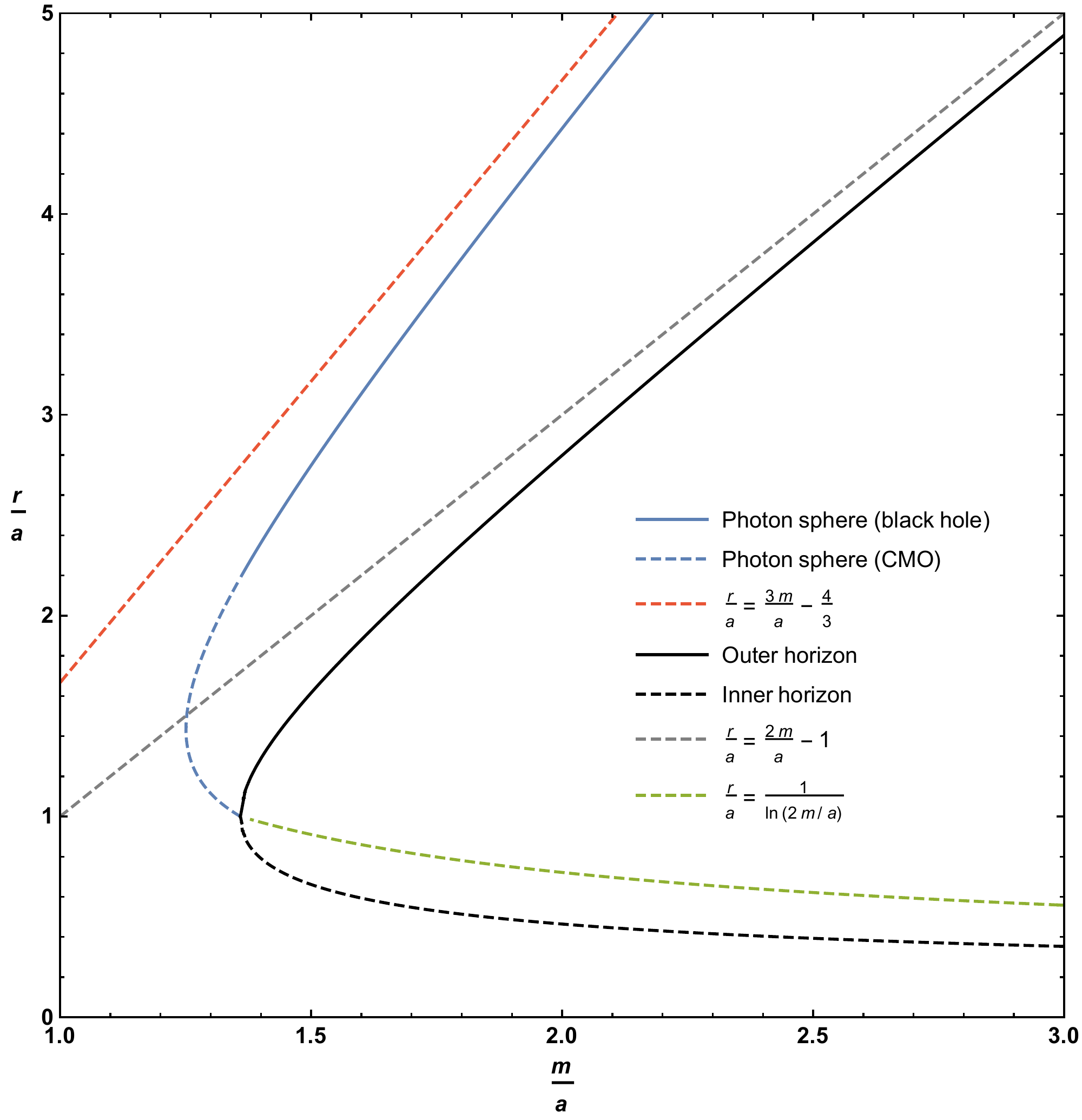}
  \caption{}
  \label{fig:iscoa}
\end{subfigure}
\caption{Location of the photon sphere, inner horizon,  and outer horizon as a function of the parameters \( a \) and \( m \). The dashed blue line represents the extension of the photon sphere to horizonless compact massive objects (CMOs), whilst the dashed red line is the asymptotic solution for small values of the parameter \( a \). (Equation \eqref{E:small-a}.) The dashed grey line is the asymptotic solution to the outer horizon for small values of \( a \). (Equation \eqref{E:outer-pert}.)  The dashed green line is the simple analytic bound and asymptotic estimate for the location of the inner horizon. (Equations \eqref{E:inner-bound} and \eqref{E:inner-asymp2}.) }
\label{fig:photonsphere}
\end{figure}

To be explicit about this, the location of a circular photon orbit, \( r_c \), is given implicitly by the equation
\begin{equation}
r_c^2 = m\,\e^{-a/r_c}(3r_c-a),
\label{eq;photonsphere}
\end{equation}
where \( a \) and \( m \) are fixed by the geometry of the spacetime.\footnote{As $a\to 0$ we have $r_c\to 3m$, as expected for Schwarzschild spacetime.}
The curve described by the loci of these circular photon orbits has been plotted in two distinct ways in Figure~\ref{fig:photonsphere}.

For clarity, defining  \( w = r_c/a \) and \( z = m/a \), we can re-write the condition for circular photon orbits as
\begin{equation}
w^2 = z \, \e^{-1/w} (3w-1);
\qquad \implies \qquad 
 z = {w^2 \; \e^{1/w}\over3w-1}.
 \label{E:z-for-photon}
\end{equation}
In Figure \ref{fig:photonsphere} we also plot the locations of both inner and outer horizons. 

The inner and outer horizons merge at $a/m=2/\e = 0.7357588824...$;
that is, at $m/a = \e/2=1.359140914...$.   For $a/m>2/\e$; that is for $m/a < \e/2$; one is dealing with a horizonless compact object and we see that there is a region where there are \emph{two} circular photon orbits. 
Note that the curve described by the loci of circular photon orbits terminates once one hits a horizon, that is, at $w=1$. Sub-horizon curves of constant $r$ are spacelike (tachyonic), and \emph{cannot} be  lightlike, so they are explicitly excluded. 
That is, photon spheres can only \emph{exist} in the region $w\in(1,\infty)$. 

Can we be more explicit about the key qualitative and quantitative features of this plot? 
Specifically, let us now analyze stability \emph{versus} instability, and find the exact location of the various turning points.

%========================================================
%========================================================

%========================================================
\subsection{Stability \emph{versus} instability for circular photon orbits} 
%========================================================

To check the \emph{stability} of these circular photon orbits we now need to investigate
\begin{equation}
V_0''(r_c) = \frac{2L^2}{r_c^7} \left[ 3r_c^3 - m\,\e^{-a/r_c}(6r_c-a)(2r_c-a) \right].
\label{E:V''}
\end{equation}

%========================================================
\subsubsection{Perturbative analysis (small $a$)} 
%========================================================

We note that determining $r_c(m,a)$ from equation (\ref{eq;photonsphere}) is  not analytically feasible, but $r_c(m,a)$ can certainly be estimated perturbatively for small $a$. 
We have
\begin{equation}
r_c(m,a) = 3m - \frac{4ma}{r_c} + \mathcal{O}(a^2) 
\quad\implies\quad 
r_c(m,a) = 3m - {4\over3} a + \mathcal{O}(a^2).
\label{E:small-a}
\end{equation}
So, for small values of \( a \), we recover the standard result for the location of the photon sphere in  Schwarzschild spacetime.

Estimating $V_0''(r_c)$ by  now substituting the approximate location of the photon sphere as \( r_c(m,a) = 3m - 4a/3 +\mathcal{O}(a^2),\) we find
\begin{equation}
V_0''(r_c(m,a)) =  -{2L^2\over81 m^4}\left(1 +  {4\over3}\, {a\over m} +\mathcal{O}(a^2)\right) .
\label{E:V''-small-a}
\end{equation}
This quantity is manifestly negative for small $a$. 
That is, (within the limits of the current small-$a$ approximation), photons are in an unstable orbit at the small-$a$ photon sphere.

%========================================================
\subsubsection{Non-perturbative analysis} 
%========================================================

However, if we rephrase the problem then we can make some much more explicit exact statements that are no longer perturbative in small $a$: Whereas determining $r_c(m,a)$ is analytically infeasible it should be noted that in contrast both $a(m,r_c) $ and $m(r_c,a)$ are easily determined analytically:
\begin{equation}
a(m,r_c) = r_c (3 - W(r_c \e^3/m)); \qquad m(r_c,a) = {r_c^2 \; \e^{a/r_c} \over(3r_c-a)}.
\label{E:non-pert}
\end{equation}
Consequently, at the peak we can write
\begin{equation}
V_0(r_c,m) = {L^2\over r_c^2} \left( 1- {2\over W(r_c \e^3/m)}\right);
\qquad
V_0(r_c,a) = {L^2\over r_c^2} \; {r_c-a\over 3r_c-a}. 
\end{equation}
Regarding stability, in the first case, substituting~(\ref{E:non-pert}\;a) into \eqref{E:V''}, we have
\begin{equation}
V_0''(r_c,m) = -{2L^2
\left( W(r_c \e^3/m)^2- W(r_c \e^3/m)-3\right)\over r_c^4  W(r_c \e^3/m)} .
\end{equation}
Using properties of the Lambert $W$ function, we quickly see that this is negative for $r_c/m >  {1\over 2} (1+\sqrt{13}) \; \e^{-5/2 +\sqrt{13}/2} = 1.146702958...$, implying instability of the circular photon orbits in this region, (and stability outside this region).

That is, on the curve of circular photon orbits, $V''(r_c)=0$ at the point
\begin{equation}
(r_c/m, a/m)_* = ( 1.146702958..., 0.7995092385...).
\end{equation}

In the second case, substituting~(\ref{E:non-pert}\;b) into \eqref{E:V''}, we have
\begin{equation}
V_0''(r_c,a) = -{2L^2\over r_c^5} \; {3r_c^2-5ar_c+a^2\over 3r_c-a}. 
\end{equation}
This will certainly be negative for $r_c/a > (5+\sqrt{13})/6 = 1.434258546...$, implying instability of the circular photon orbits in this region, (and stability outside this region).

That is, on the curve of circular photon orbits, $V''(r_c)=0$ at the point
\begin{equation}
(r_c/a,m/a)_* = ( 1.434258546..., 1.250767286...).
\end{equation}
Consequently, on the curve of circular photon orbits we have \emph{existence} and \emph{stability} in the region $w\in(1,1.434258546...)$; and \emph{existence} and \emph{instability} in the region $w\in(1.434258546...,\infty)$. 
Precisely at the point $w=1.434258546...$ the photon sphere exhibits neutral stability.

%\clearpage
%\enlargethispage{40pt}
%========================================================
\subsection{Turning points} 
%========================================================
\enlargethispage{40pt}

To evaluate the exact location of the turning points on the curve described by the loci of circular photon orbits, recall that using  \( w = r_c/a \) and \( z = m/a \) we can write this curve as 
\begin{equation}
w^2 = z \, \e^{-1/w} (3w-1) \qquad \implies \qquad z = {w^2 \e^{1/w}\over (3w-1)}.
\label{E:z-for-photon2}
\end{equation}
%\enlargethispage{20pt}
This allows us to calculate
\begin{equation}
\dv{z}{w} = \e^{1/w} \, \frac{3w^2 - 5w +1}{(3w-1)^2},
\end{equation}
which has a zero located at \( w = (5+\sqrt{13})/6 \), where we have already seen that $V_0''(r_c,a)=V_0''(w)=0$.

At this point $z$ takes on its maximum value
\begin{equation}
z = \e^{6/(5+\sqrt{13})} \frac{(5+\sqrt{13})^2}{18(3+\sqrt{13})}  =
\e^{(5-\sqrt{13})/2}\;{(2+\sqrt{13})\over 9}.
\end{equation}
Consequently, no photon sphere can exist if 
\begin{equation}
\frac{a}{m} > \e^{-(5-\sqrt{13})/2}\;(\sqrt{13}-2) = 0.7995092385...;
\label{eq;photonlimit1}
\end{equation}
or equivalently
\begin{equation}
\frac{m}{a} < \e^{(5-\sqrt{13})/2}\;{(2+\sqrt{13})\over 9} =1.250767286....
\label{eq;photonlimit2}
\end{equation}
Note that this happens when
\begin{equation}
{r_c\over m} > {1\over2} (1+\sqrt{13}) \e^{-(5-\sqrt{13})/2};
\qquad\qquad
{r_c\over a} > {5+\sqrt{13}\over 6},
\end{equation}
which was where, as we have already seen, $V_0''(r_c,m)=0$.

\begin{figure}[!htbp]
\centering
\begin{subfigure}{.5\textwidth}
  \centering
  \includegraphics[width=.95\linewidth]{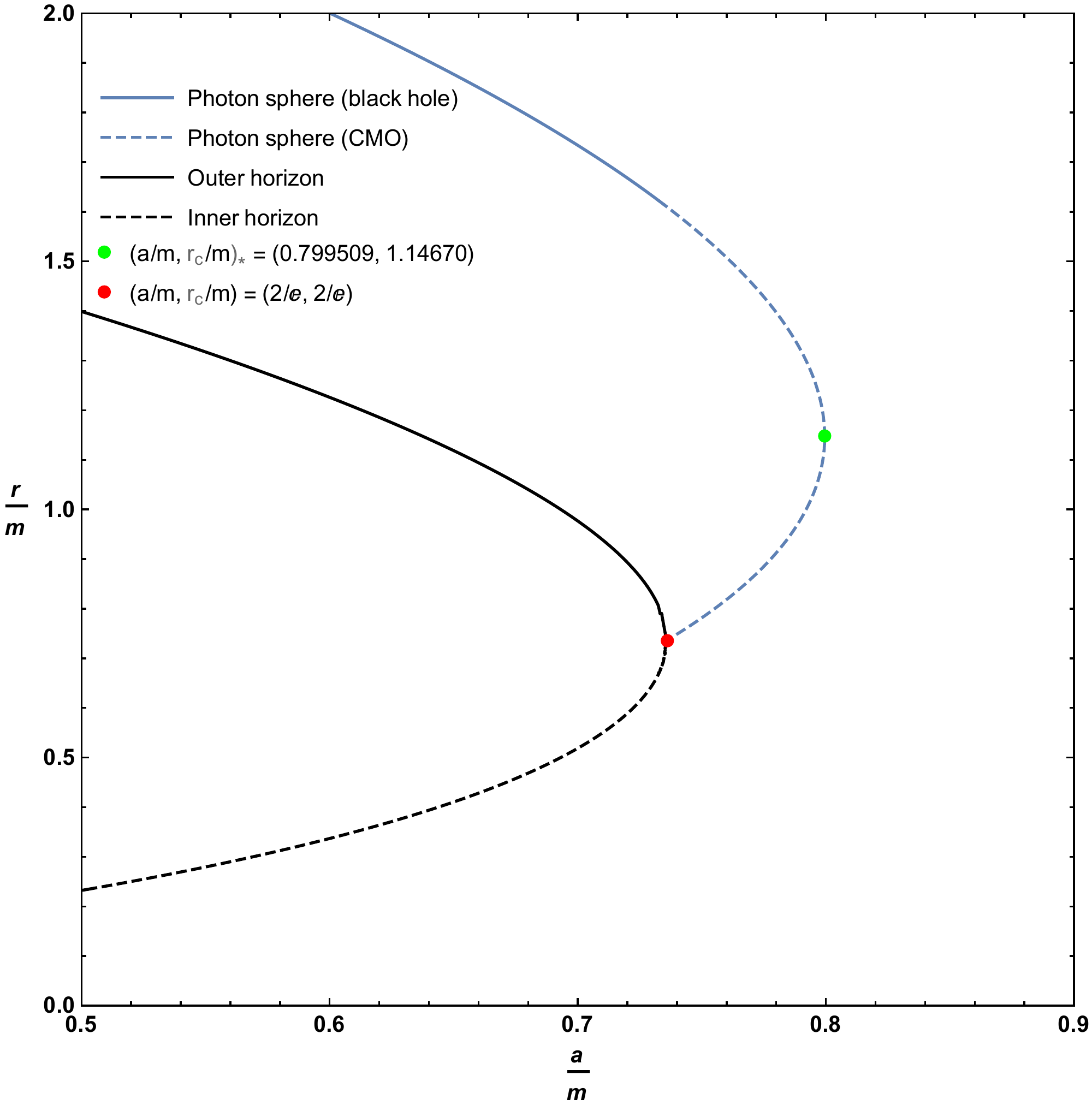}
  \caption{}
  \label{fig:iscom}
\end{subfigure}%
\begin{subfigure}{.5\textwidth}
  \centering
  \includegraphics[width=.95\linewidth]{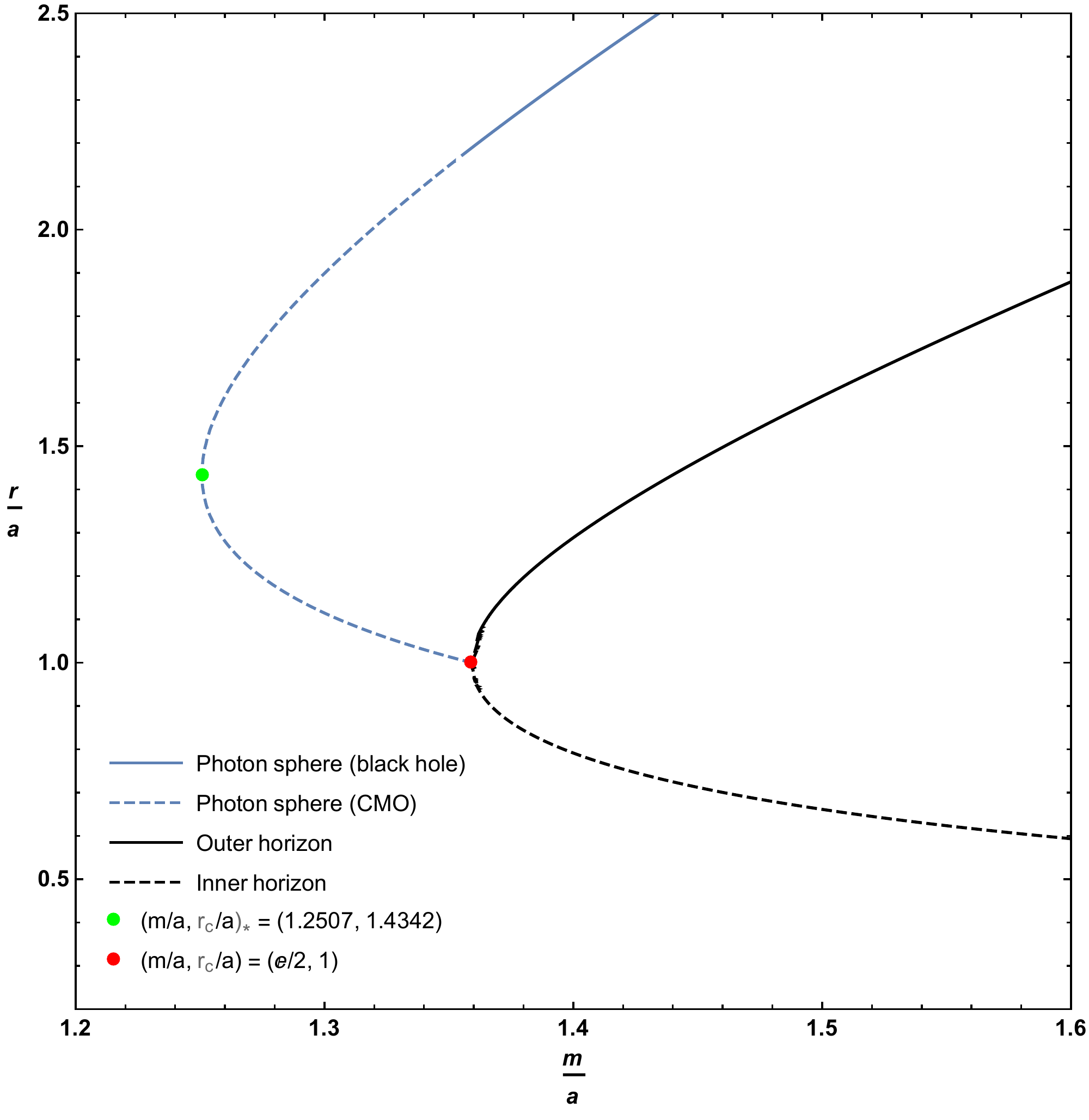}
  \caption{}
  \label{fig:iscoa}
\end{subfigure}
\caption{Zoomed in plots of the location of the photon sphere, inner horizon,  and outer horizon as a function of the parameters \( a \) and \( m \), focussing on the extremal and merger regions. The dashed blue line represents the extension of the photon sphere to horizonless compact massive objects (CMOs). Whenever the location of the photon sphere is double-valued the upper branch corresponds to an unstable photon orbit while the lower branch corresponds to a stable photon orbit.}
\label{fig:photonsphere2}
\end{figure}

%\newpage
As can be seen, originally from Figure~\ref{fig:photonsphere}, and now in more detail in the zoomed-in plot in Figure~\ref{fig:photonsphere2}, for horizonless compact massive objects there is a region where there are two possible locations for the photon sphere for fixed values of \( m \) and \( a \). Furthermore when this happens it is the upper branch that corresponds to an unstable photon orbit, while the lower branch is a stable photon orbit.
\enlargethispage{40pt}

%========================================================
% ZOOMED
%========================================================
%========================================================

\newpage
%===============================================
\section{Timelike circular orbits}
%===============================================
\def\ISCO{{\hbox{\tiny ISCO}}}
\def\ESCO{{\hbox{\tiny ESCO}}}
%===============================================

Let us first check the \emph{existence}, and then the \emph{stability}, of timelike circular orbits.
Even in Schwarzschild spacetime ($a\to0$) this is not entirely trivial: Timelike circular orbits \emph{exist} for all $r_c\in (3m,\infty)$; they are unstable for $r_c\in (3m,6m)$,
exhibit neutral stability at $r_c=6m$, and are stable for $r_c\in (6m,\infty)$. 
Once the parameter $a$ is non-zero the situation is much more complex.

%========================================================
\subsection{Existence of circular timelike orbits} 
%========================================================

For timelike trajectories, the effective potential is given by
\begin{equation}
V_{-1}(r) = \left(1-\frac{2m\,\e^{-a/r}}{r}\right)\left(1+\frac{L^2}{r^2}\right),
\end{equation}
and so the locations of the circular orbits can be found from
\begin{equation}
V_{-1}'(r_c) = -\frac{2}{r_c^5}\left\{ L^2r_c^2+m\,\e^{-a/r_c} [ a(L^2+r_c^2)-r_c(3L^2+r_c^2) ] \right\} = 0.
\end{equation}
That is, all timelike circular orbits (there will be infinitely many of them) must satisfy
\begin{equation}
 L^2r_c^2+m\,\e^{-a/r_c} [ a(L^2+r_c^2)-r_c(3L^2+r_c^2) ]  = 0.
\end{equation}

This is not analytically solvable for \( r_c(L,m,a) \), however we \emph{can} solve for the required angular momentum $L_c(r_c,m,a)$ of these circular orbits:
\begin{equation}
L_c(r_c,m,a)^2 = {\frac{r_c^2 \, m(r_c-a)}{ma-3mr_c+r_c^2\,\e^{a/r_c}}}.
\label{E:Lsq}
\end{equation}
Physically we must demand $0\leq L_c^2 <\infty$, so the boundaries for the \emph{existence} region of circular orbits (whether stable or unstable) are given by
\begin{equation}
r_c = a; \qquad\qquad {ma-3mr_c+r_c^2\,\e^{a/r_c}}=0.
\end{equation}
The first of these conditions $r_c=a$, comes from the fact that in this spacetime gravity is effectively repulsive for $r<a$. 
Remember that $g_{tt} = -(1-2m\e^{-a/r}/r)$, and that the pseudo-force due to gravity depends on $\partial_r g_{tt}$. Specifically
\begin{equation}
\partial_r g_{tt} = - {2m\over r^2} \; \e^{-a/r} \; \left(1-{a\over r}\right),
\end{equation}
and this changes sign at $r=a$. 
So for $r>a$ gravity attracts you to the centre, but for $r<a$ gravity repels you from the centre.

And if gravity repels you, there is no way to counter-balance it with a centrifugal pseudo-force, and so there is simply no way to get a circular orbit, regardless of whether it be stable or unstable.
Precisely at $r=a$ there are stable ``orbits'' where the test particle just sits there, with zero angular momentum, no sideways motion required.
Since by construction $r_c>r_{H^+} \geq a$, this constraint is relevant only for horizonless CMOs. 

 The second of these conditions is exactly the location of the photon orbits considered in the previous sub-section. (Physically what is going on is this: At large distances it is easy to put a massive particle into a circular orbit with $L_c\propto\sqrt{mr_c}$. As one moves inwards and approaches the photon orbit, the massive particle must move more and more rapidly, and the angular momentum per unit mass must diverge when a particle with nonzero invariant mass tries to orbit at the photon orbit.)
 
\begin{figure}[!htbp]
\centering
\begin{subfigure}{.5\textwidth}
  \centering
  \includegraphics[width=.95\linewidth]{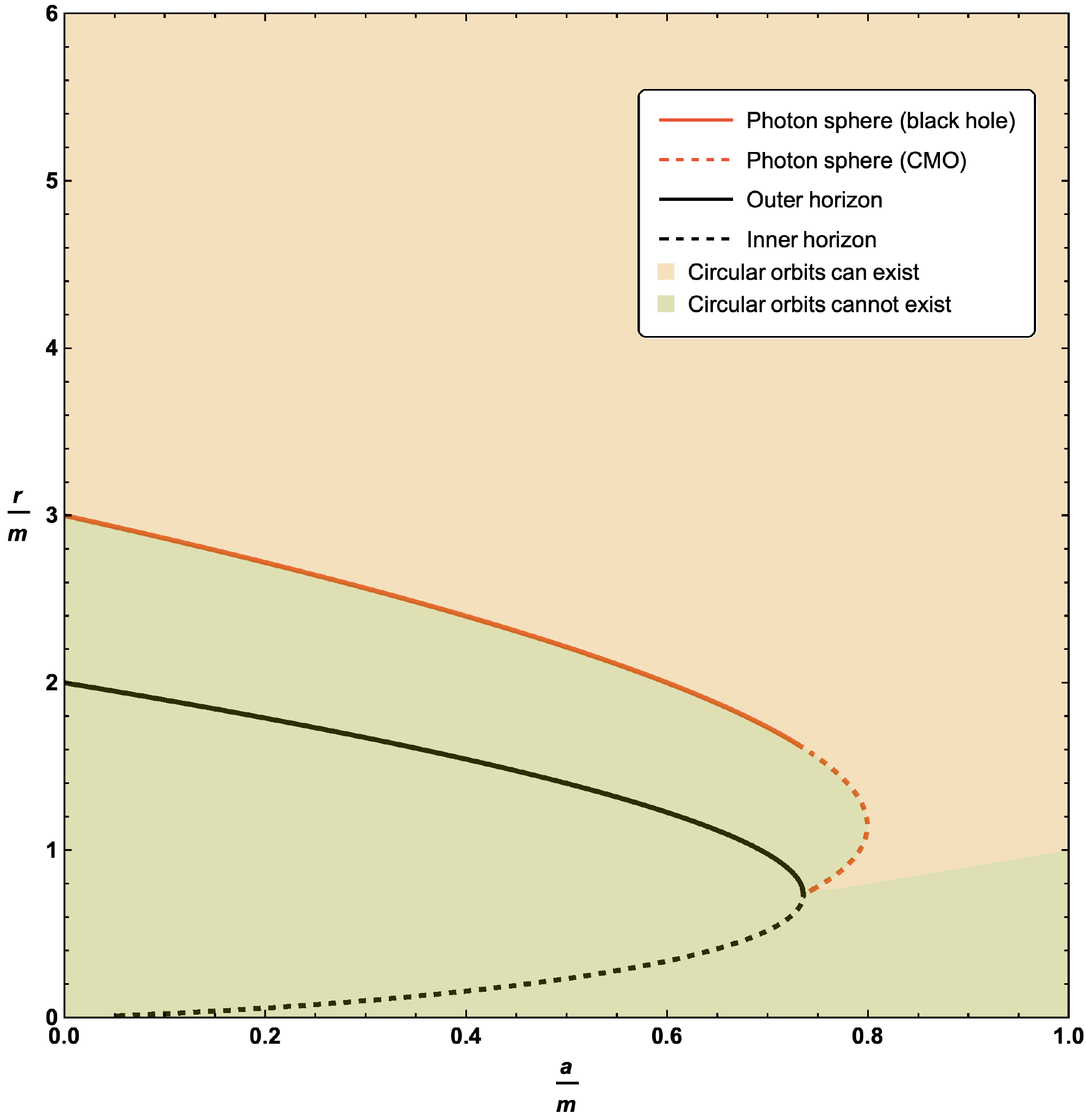}
  \caption{}
  \label{fig:iscom}
\end{subfigure}%
\begin{subfigure}{.5\textwidth}
  \centering
  \includegraphics[width=.95\linewidth]{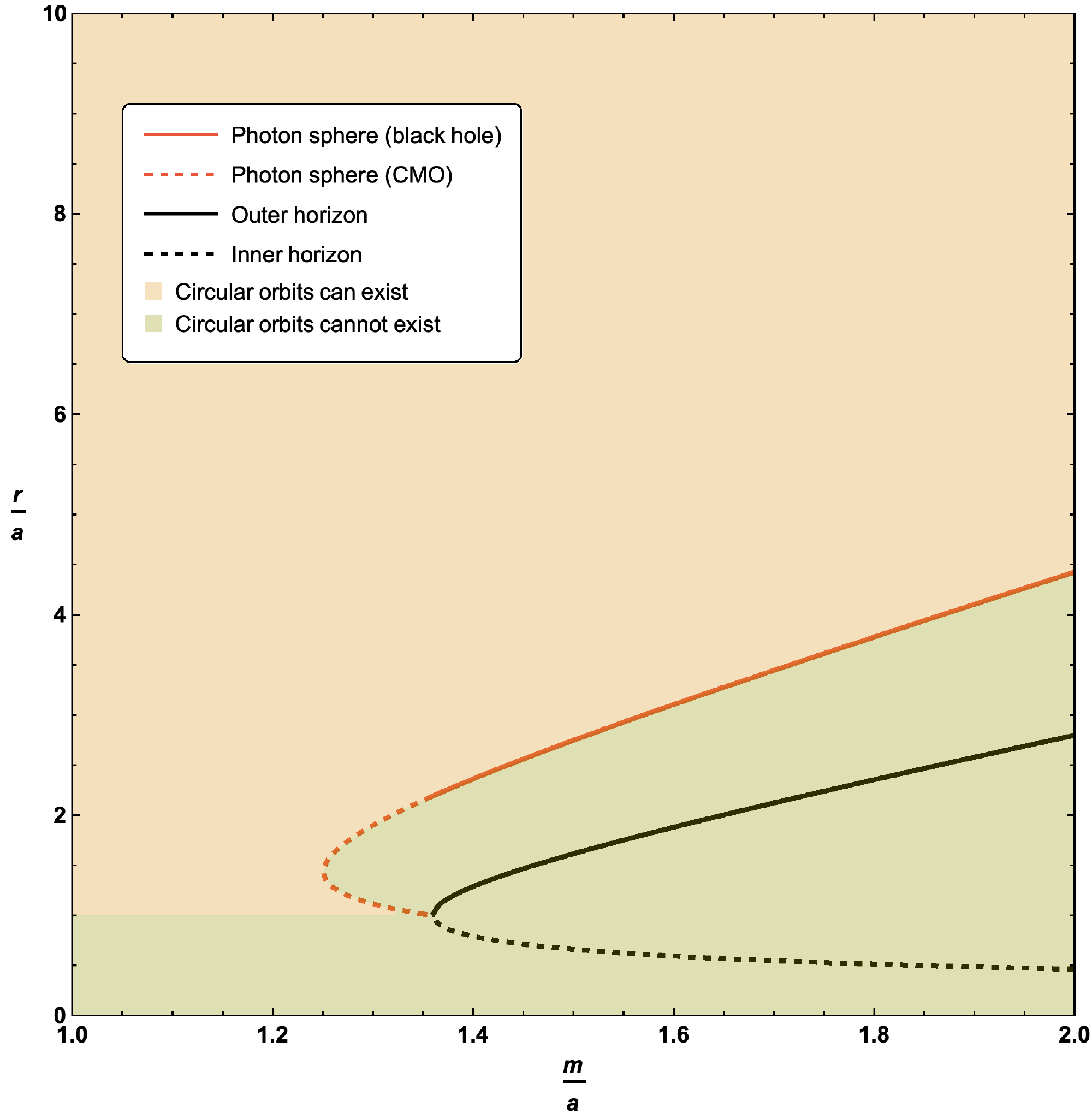}
  \caption{}
  \label{fig:iscoa}
\end{subfigure}
\caption{Locations of the \emph{existence} region for timelike circular orbits in terms of the circular null geodesics, outer horizon, and inner horizon for various values of the parameters \( a \) and \( m \). }
\label{fig:isco-existence}
\end{figure}

 Thus the existence region (rather than just its boundary) for timelike circular orbits is therefore:
 \begin{equation}
r_c > a; \qquad\qquad {ma-3mr_c+r_c^2\,\e^{a/r_c}}>0.
\end{equation}
See Figure~\ref{fig:isco-existence}.
 %========================================================
 % EXISTENCE
 %========================================================
%========================================================
 
%========================================================
\subsection{Stability \emph{versus} instability for  circular timelike orbits} 
%========================================================

Now consider the general expression
\begin{equation}
V_{-1}''(r) = {6L^2 r^3- 2m(2r^4-4ar^3+(12L^2+a^2)r^2-8L^2ar+L^2a^2)\e^{-a/r}\over r^7},
\end{equation}
and substitute the known value of $L\to L_c(r_c)$ for circular orbits, see \eqref{E:Lsq}. Then
\begin{equation}
V_{-1}''(r_c) = -{2m e^{-a/r_c}(2m(3r_c^2-3ar_c+a^2) \e^{-a/r_c} -r_c(r_c^2+ar_c-a^2))\over
(r_c^2 - m(3r_c-a) \e^{-a/r_c})r^4}.
\end{equation}
Note that $V_{-1}''(r_c) \to\infty$ at the photon orbit, (where the denominator has a zero).

To locate the \emph{boundary} of the region of \emph{stable} circular orbits, the ESCO (extremal stable circular orbit),  we now need to set $V_{-1}''(r_c)=0$, leading to the equation
\begin{equation}
2m(3r_c^2-3ar_c+a^2) \e^{-a/r_c} = r_c(r_c^2+ar_c-a^2).
\label{E:stable}
\end{equation}
We note that locating this boundary is equivalent to extremizing $L_c(r_c)$. 
To see this, consider the quantity $[V_{-1}'(L(r),r)]=0$ and differentiate:
\begin{equation}
{\dd{} [V_{-1}'(L(r),r)]\over \dd r }  = 
\left.{\partial V_{-1}'(L,r)\over\partial L}\right|_{L=L(r)} \times {\dd L(r)\over \dd r}
+ \left.V''_{-1}(L,r)\right|_{L=L(r)}.
\end{equation}
This implies
\begin{equation}
0  = 
\left.{\partial V_{-1}'(L,r)\over\partial L}\right|_{L=L(r)} \times {\dd L(r)\over \dd r}
\;+\; \left.V_{-1}''(L,r)\right|_{L=L(r)}.
\end{equation}
Thence
\begin{equation}
 \left.V_{-1}''(L,r)\right|_{L=L(r)}  = - \left.{\partial V_{-1}'(L,r)\over\partial L}\right|_{L=L(r)} 
 \times {\dd L(r)\over \dd r}.
 \end{equation}
But it is easily checked that ${\partial V_{-1}'(L,r)/\partial L}$ is non-zero outside the photon sphere, (that is, in the existence region for circular timelike geodesics).
Thence:
\begin{equation}
 \left.V_{-1}''(L,r)\right|_{L=L(r)}  = 0
 \qquad \Longleftrightarrow  \qquad
 {\dd L(r)\over \dd r}=0.
 \end{equation}
So one might a well extremize $L^2_c(r_c)$, as in equation (\ref{E:Lsq}), and one again finds equation (\ref{E:stable}). 

Defining  \( w = r_c/a \) and \( z = m/a \) the curve describing the boundary of the region of stable timelike circular orbits can be rewritten as
\begin{equation}
\label{E:dd}
2z (3w^2-3w+1)\e^{-1/w} = w(w^2+w-1).
\end{equation}

\begin{figure}[!htbp]
\centering
\begin{subfigure}{.5\textwidth}
  \centering
  \includegraphics[width=.95\linewidth]{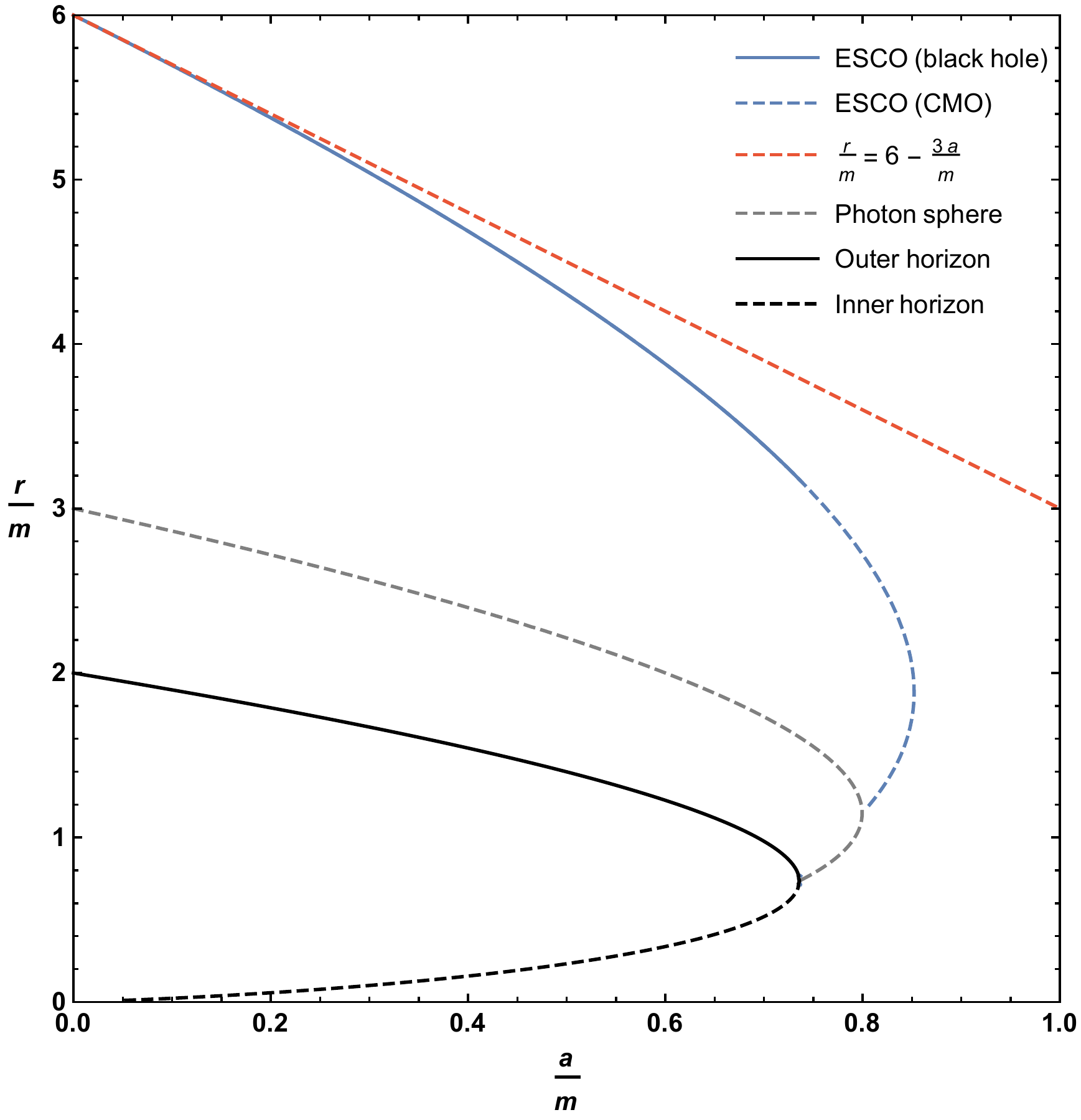}
  \caption{}
  \label{fig:iscom}
\end{subfigure}%
\begin{subfigure}{.5\textwidth}
  \centering
  \includegraphics[width=.95\linewidth]{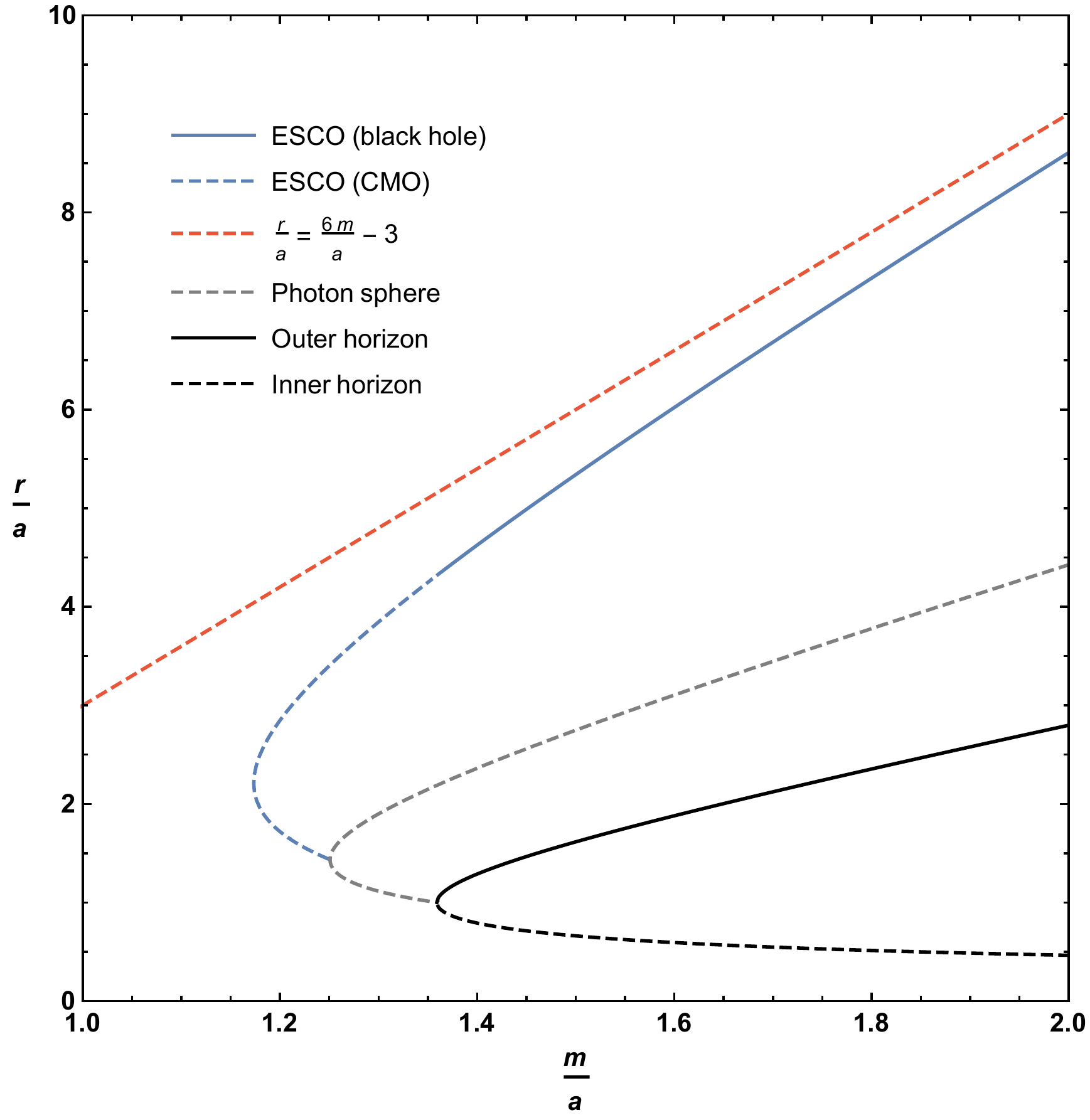}
  \caption{}
  \label{fig:iscoa}
\end{subfigure}
\caption{Locations of the ESCO, photon sphere, outer horizon, and inner horizon for various values of the parameters \( a \) and \( m \).
The dashed blue line represents the extension of the ESCO to CMOs. The dashed red curves in sub-figure (a) and (b) is the asymptotic location of the ISCO for small values of \( a \) (approaching the Schwarzschild solution).}
\label{fig:esco}
\end{figure}

\begin{figure}[!htbp]
\centering
\begin{subfigure}{.5\textwidth}
  \centering
  \includegraphics[width=.95\linewidth]{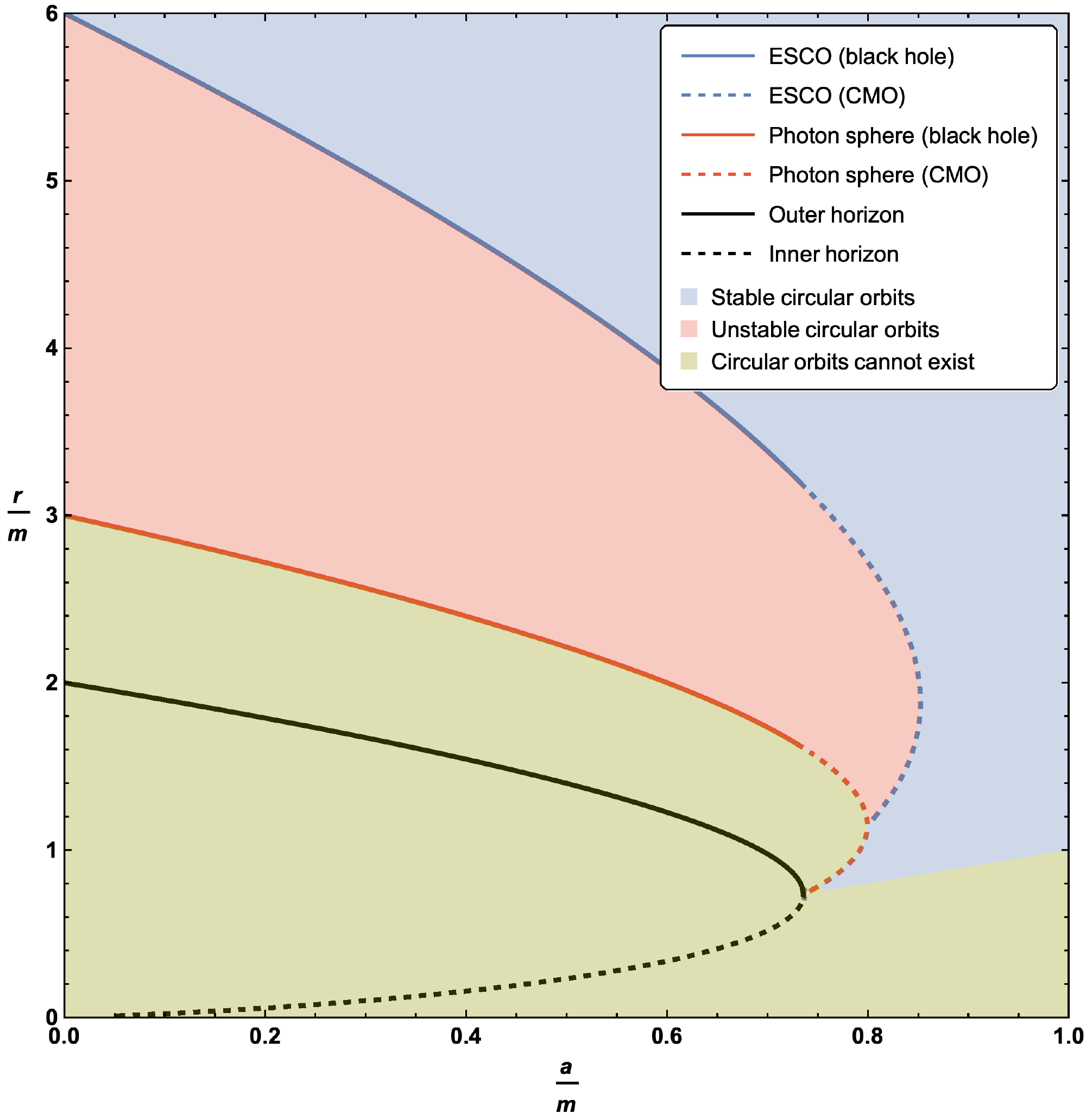}
  \caption{}
  \label{fig:iscom}
\end{subfigure}%
\begin{subfigure}{.5\textwidth}
  \centering
  \includegraphics[width=.95\linewidth]{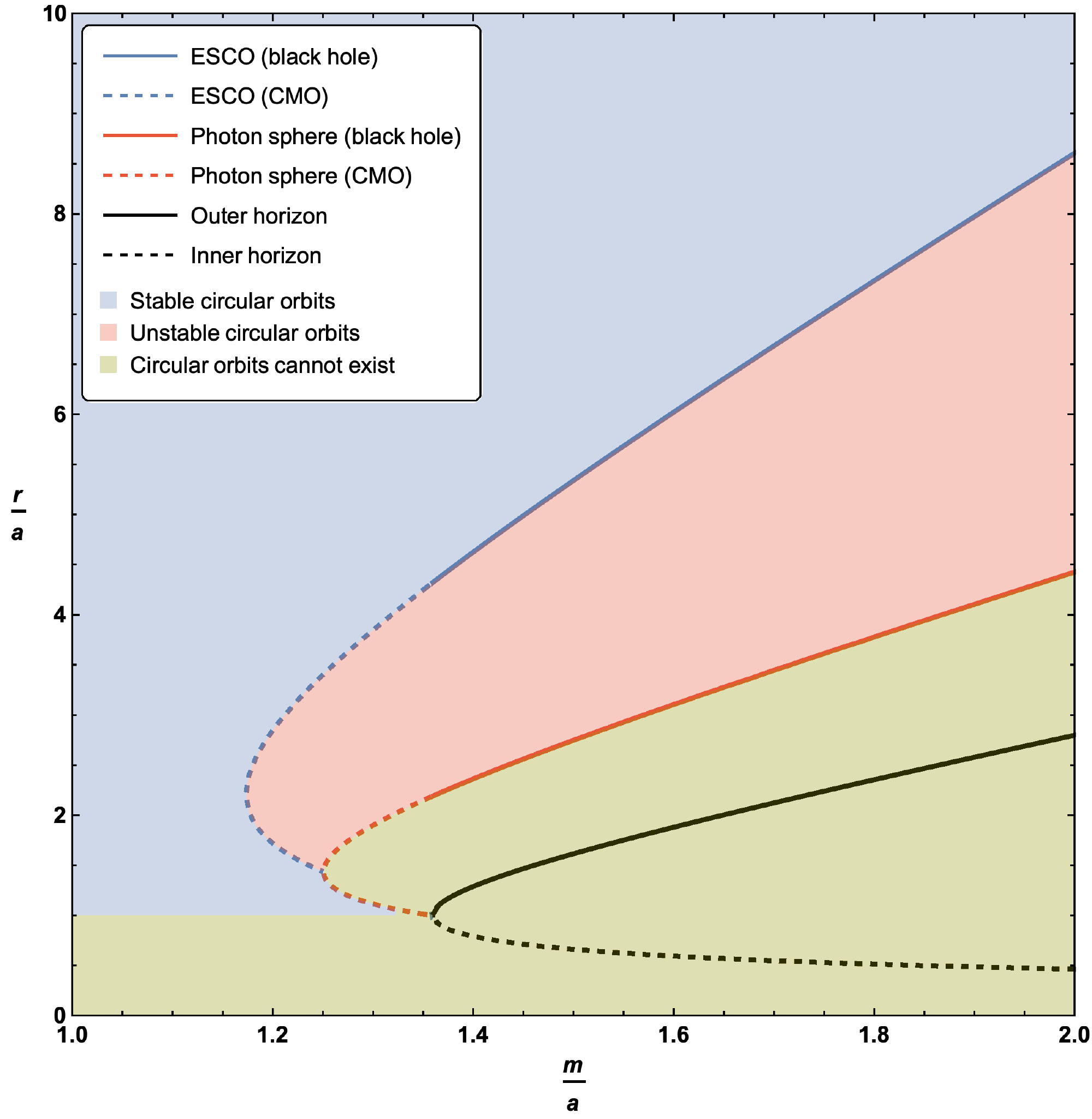}
  \caption{}
  \label{fig:iscoa}
\end{subfigure}
\caption{Locations of the ESCO, photon sphere, outer horizon, and inner horizon for various values of the parameters \( a \) and \( m \).
The dashed blue line represents the extension of the ESCO to CMOs. 
The dashed red line represents the extension of the photon sphere to CMOs.
The blue region denotes stable timelike circular orbits, while the red region denotes unstable timelike circular orbits, and the green region denotes the non-existence of timelike circular orbits. 
Where the ESCO is single-valued it is an ISCO. Where the ESCO is double-valued the upper branch is an ISCO and the lower branch is an OSCO (outermost stable circular orbit). 
}
\label{fig:esco-2}
\end{figure}

Plots of the boundary implied by equation \eqref{E:stable}, or equivalently \eqref{E:dd}, can be seen in Figure~\ref{fig:esco}.
As for the photon sphere, we have the interesting result that the extension of the ESCO to horizonless compact massive objects results in up to two possible ESCO locations for fixed values of \( a \) and \( m \). 
Perhaps unexpectedly, the curve of ESCOs does not terminate at the horizon --- it terminates once it hits the curve of circular photon orbits at a very special point.
Let us now turn to the detailed analysis of both the qualitative behaviour and the various turning points presented in Figures~\ref{fig:esco} and~\ref{fig:esco-2}.
Note that where the ESCO is single-valued it is an ISCO (innermost stable circular orbit). Where the ESCO is double-valued the upper branch is an ISCO and the lower branch is an OSCO (outermost stable circular orbit)~\cite{OSCO}. 

%========================================================
% ESCO --- no regions
%========================================================
%========================================================

%========================================================
% ESCO --- all regions
%========================================================
%========================================================

%========================================================
\subsubsection{Perturbative analysis (small $a$)} 
%========================================================

Let us first investigate the existence region perturbatively for small $a$. We have
\begin{equation}
L_c(r_c,m,a)^2 = {mr_c^2\over r_c-3m} -{2mr_c(r_c-m)\over(r_c-3m)^2} \; a + \mathcal{O}(a^2).
\end{equation}
\leftline{Note that this approximation diverges at the Schwarzschild photon sphere $r=3m$.}
So for small $a$ the boundary for the region of \emph{existence} of timelike circular orbits is still  $r=3m$.

\newpage
Now investigate the \emph{stability} region perturbatively for small $a$. Rearranging  
equation (\ref{E:stable}) we see
\begin{equation}
r_c = {6m(r_c^2-ar_c+a^2/3) \e^{-a/r_c}\over r_c^2+ar_c-a^2} 
= 6m\left( 1-{3a\over r_c} +\O(a^2) \right).
\end{equation}
Thence
\begin{equation}
r_c = 6m -3a +\O(a^2). 
\end{equation}
Which sensibly reproduces the Schwarzschild ISCO to lowest order in $a$, and explains the asymptote in Figure~\ref{fig:esco} (b).

Furthermore, for small $a$, substituting $L_c(r_c)$ into $V''_{-1}(L,r_c)$ and expanding
\begin{equation}
V_{-1}''(r_c) = {2m(r_c-6m)\over r_c^3(r_c-3m)} 
+  {4m^2(7r_c-15m)\over r^4(r_c-3m)^2} \; a  +\O(a^2)
\end{equation}
Demanding that this quantity be zero self-consistently yields $r_c = 6m -3a +\O(a^2)$. 

%========================================================
\subsubsection{Non-perturbative analysis} 
%========================================================

\begin{figure}[h]
\centering
\begin{subfigure}{.5\textwidth}
  \centering
  \includegraphics[width=.95\linewidth]{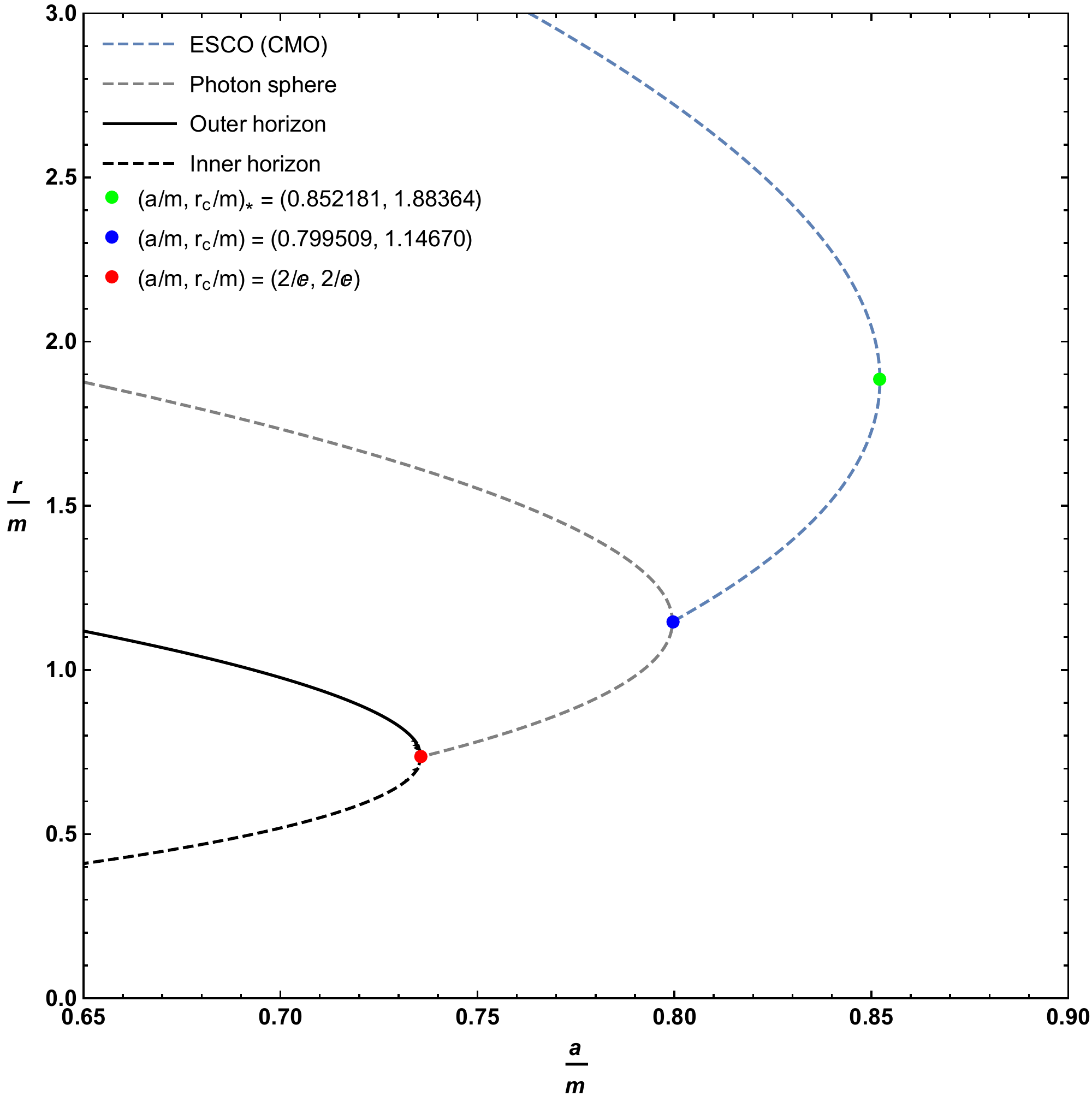}
  \caption{}
  \label{fig:iscom}
\end{subfigure}%
\begin{subfigure}{.5\textwidth}
  \centering
  \includegraphics[width=.95\linewidth]{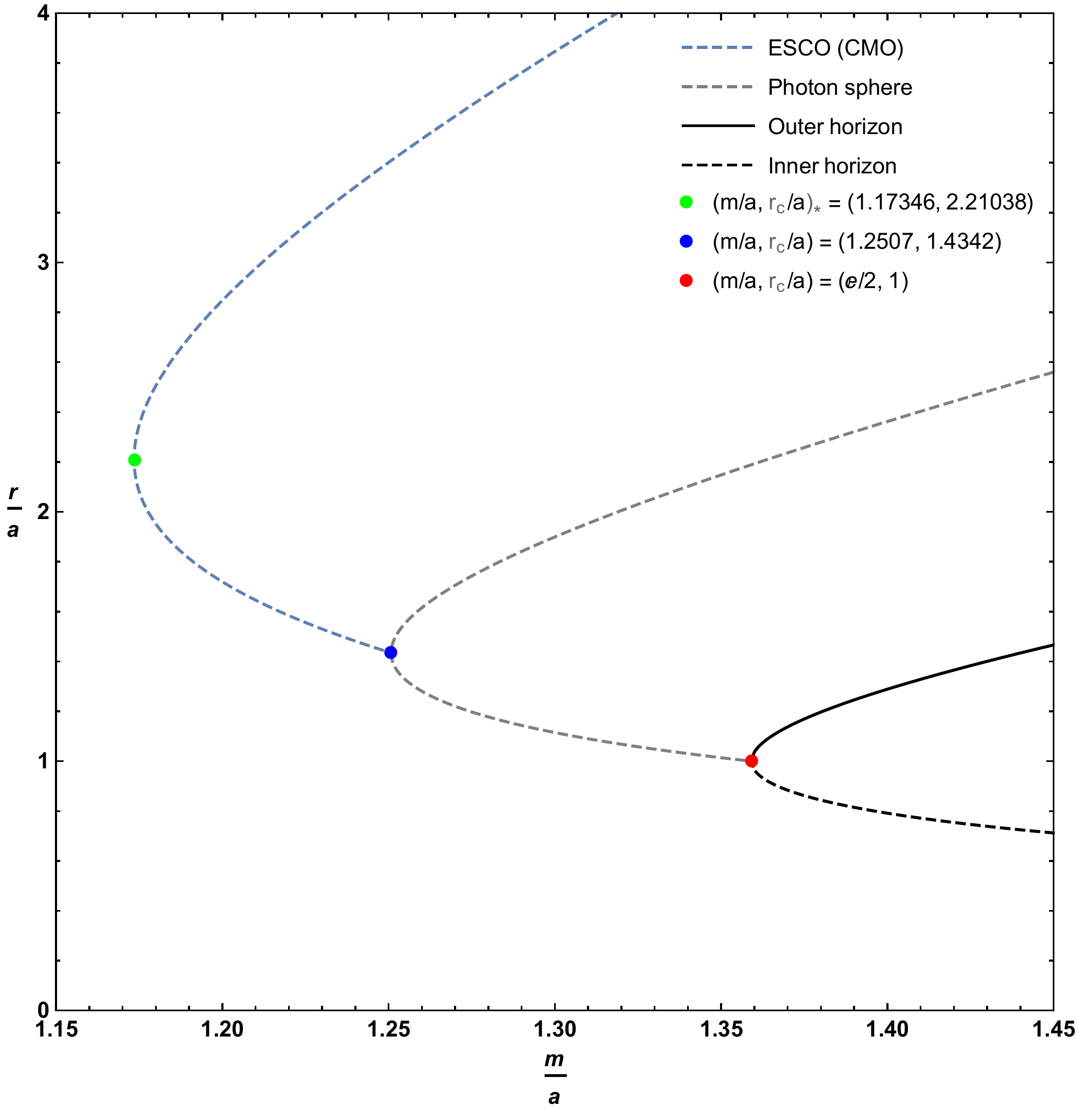}
  \caption{}
  \label{fig:iscoa}
\end{subfigure}
\caption{Zoomed in plot of the locations of the ESCO, outer horizon, and inner horizon for various values of the parameters \( a \) and \( m \), focussing on the turning points.  The dashed blue line represents the extension of the ESCO to CMOs. Where the ESCO is single-valued it is an ISCO. Where the ESCO is double-valued the upper branch is an ISCO and the lower branch is an OSCO.}
\label{fig:esco-3}
\end{figure}

We have already seen that, defining  \( w = r_c/a \) and \( z = m/a \), the curve describing the boundary of the region of stable timelike circular orbits can be rewritten as
\begin{equation}
\label{E:dd3}
2z (3w^2-3w+1)\e^{-1/w} = w(w^2+w-1).
\end{equation}
Thence
\begin{equation}
%\label{E:dd4}
    z = {w(w^2+w-1)\e^{1/w}\over 2(3w^2-3w+1)}.
    \label{E:z-for-esco}
\end{equation}
Let us look for the turning points of $z(w)$. The derivative is
\begin{equation}
{\dd z\over\dd w} = {(w-1)(3w^4-6w^3-3w^2+4w-1)\e^{1/w} \over 2w(3w^2-3w+1)^2}.
\end{equation}
There is one obvious local extrema at $w=1$, corresponding to $z=\e/2$. Physically this corresponds to the point where inner and outer horizon merge and become extremal --- but from inspection of Figure~\ref{fig:esco}, the descriptive plots of Figure~\ref{fig:esco-2}, and the zoomed-in plots of Figure~\ref{fig:esco-3}, we see that the curve of ESCOs hits the photon orbit (and becomes unphysical) before getting to this point.
In terms of the variables used when plotting Figures~\ref{fig:esco}--\ref{fig:esco-3} this unphysical (from the point of view of ESCOs) point corresponds to
\begin{equation}
(r_c/a, m/a)_* = (1, \e/2) \qquad (r_c/m, a/m)_* = (2/\e, 2/\e). 
\end{equation}

The other local extrema is located at the only physical root of the quartic polynomial
\begin{equation}
3w^4-6w^3-3w^2+4w-1 =0.
\end{equation}
While this can be solved analytically, the results are too messy to be enlightening and so we resort to numerics. 
Two roots are complex, one is negative, the only physical root is $w= 2.210375896...$, corresponding to $z=1.173459017...$. 
Physically this implies that the ESCO curve should exhibit a non-trivial local extremum --- and from inspection of Figure~\ref{fig:esco} we see that the curve of ESCOs does indeed have a local extremum at this point. In terms of the variables used when plotting Figure~\ref{fig:esco} this extremal point corresponds to
\begin{equation}
(r_c/a, m/a)_* = (2.210375896, 1.173459017), 
\end{equation}
and
\begin{equation}
(r_c/m, a/m)_* = (1.883641323, 0.8521814444). 
\end{equation}

\newpage
%========================================================
\subsection{Intersection of ESCO and photon sphere } 
%========================================================
\enlargethispage{40pt}

We can rewrite the curve for the loci of the photon spheres \eqref{E:z-for-photon} as 
\begin{equation}
\e^{-1/w} z = {w^2\over (3w-1)}.
\end{equation}
Similarly, for the loci of ESCOs rewrite \eqref{E:z-for-esco} as
\begin{equation}
\e^{-1/w} z = {w(w^2+w-1)\over 2 (3w^2-3w+1)}.
\end{equation}
These curves cross at
\begin{equation}
{w\over (3w-1)} = {(w^2+w-1)\over 2 (3w^2-3w+1)}.
\end{equation}
That is, at
\begin{equation}
(w-1)(3w^2-5w+1)=0,
\end{equation}
with explicit roots at
\begin{equation}
1, \quad {5\pm\sqrt{13}\over6}.
\end{equation}
The physically relevant root is  $w = {5+\sqrt{13}\over6}= 1.434258546...$, which was where we previously determined that  the photon sphere became stable, and at the point where the curve of photon spheres maximized the value of $z=m/a$.

%========================================================
% ZOOMED
%========================================================
%========================================================

%========================================================
\subsection{Explicit result for the angular momentum} 
%========================================================

We can rewrite the curve for the angular momentum \eqref{E:Lsq} as 
\begin{equation}
L_c^2 = 
a^2  \left( \e^{-1/w} z \; w^2 (w-1) \over  w^2 - \e^{-1/w} z (3w-1)\right) .
\end{equation}
Similarly, for the loci of ESCOs we can rewrite \eqref{E:z-for-esco} as
\begin{equation}
\e^{-1/w} z = {w(w^2+w-1)\over 2 (3w^2-3w+1)}.
\end{equation}
We then substitute this into back into $L_c$:
\begin{equation}
L_c^2 = a^2\; {w^2(w^2+w-1)\over 3w^2-5w+1}.
\end{equation}
This has a pole at $w = {5+\sqrt{13}\over6}= 1.434258546...$, 
and is then positive and finite for all $w >{5+\sqrt{13}\over6}$.  
(Of course the point $w = {5+\sqrt{13}\over6}$ on the ESCO curve is exactly where the ESCO curve hits the photon curve, so we would expect the angular momentum to go to infinity there.)
Asymptotically for large $r$ (large $w=r_c/a$) we have $L_c^2 \sim a^2 w^2/3$ and $m/a = z \sim w/6$, so $L_c^2 \sim 2 m r_c$ as expected from the large-distance Newtonian limit. 

%========================================================
\subsection{Summary} 
%========================================================

Overall, we see that the boundary of the stability region for timelike circular orbits 
is rather complicated. In terms of the variable $w=r_c/a$:
\begin{itemize}
\item For $w\in ({5+\sqrt{13}\over6}, \infty)$ we have an ESCO. \\[3pt]
This ESCO then subdivides as follows:
\begin{itemize}
\item For $w \in( 2.210375896, \infty)$ we have an ISCO.
\item For $w \in ({5+\sqrt{13}\over6}, 2.210375896)$ we have an OSCO.
\end{itemize}
\item For $w\in (1, {5+\sqrt{13}\over6})$ the stability region is bounded by a stable photon orbit.
\item The line $w=1$ bounds the stability and existence region for timelike circular orbits from below.
\end{itemize}
This is considerably more complicated than might reasonably have been expected.

\newpage
%========================================================
\section{Conclusions}
%========================================================

In this work we have investigated astrophysically observable quantities of a specific novel regular black hole model based on an asymptotically Minkowski core~\cite{asymptot-mink-core,Berry:2020}: Specifically we have investigated the photon sphere and ESCO.
The spacetime under consideration is an example of a black hole mimicker.
For the regular black hole model, both the photon sphere and the ESCO exist and are located outside of the outer horizon, and so (at least in theory) could be astrophysically observable.
The analysis of the photon sphere and ESCO was extended to horizonless compact massive objects, leading to the surprising results that for fixed values of \( m \) and \( a \), up to two possible photon sphere and up to two possible ESCO locations exist in our model spacetime; and that the very existence of the photon sphere and ESCO depends explicitly on the ratio \( a/m \). Somewhat unexpectedly, due to the effectively repulsive nature of gravity in the region near the core, we have found some situations in which the photon orbits are stable, and some situations where the ESCOs are OSCOs rather than ISCOs. 
There is a rich phenomenology here that is significantly more complex than for the Schwarzschild spacetime.

%=====================================================
\section*{Acknowledgements}
%=====================================================

TB was supported by a Victoria University of Wellington MSc scholarship, and was also indirectly supported by the Marsden Fund, via a grant administered by the Royal Society of New Zealand. \\
AS acknowledges financial support via a PhD Doctoral Scholarship provided by Victoria University of Wellington.
AS is also indirectly supported by the Marsden fund, via a grant administered by the Royal Society of New Zealand. \\
MV was directly supported by the Marsden Fund, via a grant administered by the Royal Society of New Zealand.

%=====================================================
%===================================================== 

%=====================================================
%===================================================== 

%==================================================================

\begin{thebibliography}{99}
%=====================================================
%===================================================== 

\bibitem{schwarzschild-original}
K. Schwarzschild,
``\"Uber das Gravitationsfeld eines Massenpunktes nach der Einsteinschen Theorie'',
Sitzungsberichte der K\"oniglich Preussischen Akademie der Wissenschaften \textbf{7} (1916) 189.
\href{https://tinyurl.com/y99c33n5}{Free online version}.

%%%%%%%%%%%%%%%%%%%%%%%%%%%%%%%%%%%%%%
\bibitem{reissner-original}
H. Reissner,
``\"Uber die Eigengravitation des elektrischen Feldes nach der Einsteinschen Theorie'',
Annalen der Physik \textbf{50} (1916) 106. 
\href{https://tinyurl.com/ycyh7kat}{Free online version}.

\bibitem{weyl-original}
H. Weyl,
``Zur Gravitationstheorie'',
Annalen der Physik \textbf{54} (1917) 117. \\
\href{https://tinyurl.com/y9xw8dgz}{Free online version}.

\newpage
\bibitem{nordstrom-original}
G. Nordstr\"om,
``On the Energy of the Gravitational Field in Einstein's Theory'',
Verhandl. Koninkl. Ned. Akad. Wetenschap., \\
Afdel. Natuurk., Amsterdam \textbf{24} (1918) 1201.

\bibitem{kerr-original}
R. Kerr,
``Gravitational Field of a Spinning Mass as an Example of Algebraically Special Metrics'',
Phys. Rev. Lett. \textbf{11} (1963) 237.

\bibitem{kerr-newmann}
E. Newmann, E. Couch, K. Chinnapared, A. Exton, A. Prakash and R. Torrence,
``Metric of a Rotating, Charged Mass'',
J. Math. Phys. \textbf{6} (1965) 918.

\bibitem{kerr-schild}
R. Kerr and A. Schild,
``Republication of: A new class of vacuum solutions of the Einstein field equations'',
Gen. Rel. and Grav. \textbf{41} (2009),  2485. \\
(Original paper published 1965.)

\bibitem{kerr-intro}
M.~Visser,
``The Kerr spacetime: A brief introduction'',
[\href{https://arxiv.org/abs/0706.0622}{arXiv:0706.0622} [gr-qc]].
Published in \cite{kerr-book}.
%94 citations counted in INSPIRE as of 13 Jun 2020

\bibitem{kerr-book}
D.~L.~Wiltshire, M.~Visser and S.~M.~Scott,\\
\emph{The Kerr spacetime: Rotating black holes in general relativity},\\
(Cambridge University Press, Cambridge, 2009).
%31 citations counted in INSPIRE as of 13 Jun 2020

\bibitem{Baines:2020}
J.~Baines, T.~Berry, A.~Simpson and M.~Visser,\\
``Unit-lapse versions of the Kerr spacetime'',
[\href{https://arxiv.org/abs/2008.03817}{arXiv:2008.03817} [gr-qc]].
%0 citations counted in INSPIRE as of 20 Aug 2020

\bibitem{Lense-Thirring}
J.~Baines, T.~Berry, A.~Simpson and M.~Visser,\\
``Painleve-Gullstrand form of the Lense-Thirring spacetime,''\\{}
[\href{https://arxiv.org/abs/2006.14258}{arXiv:2006.14258} [gr-qc]].
%1 citations counted in INSPIRE as of 25 Aug 2020

%%%%%%%%%%%%%%%%%%%%%%%%%%%%%%%%%%%%%%%%%%%%%%%%
\bibitem{vaidya-original1}
P. C. Vaidya,
``The External Field of a Radiating Star in General Relativity'',
Current Sci. (India) \textbf{12} (1943) 183.

\bibitem{vaidya-original2}
P. C. Vaidya,
``The external field of a radiating star'',\\
Proc. Indian Acad. Sci. \textbf{33} (1951) 264.

\bibitem{vaidya-original3}
P. C. Vaidya,
``Nonstatic solutions of Einstein's field equations for spheres of fluids radiating energy''
Phys. Rev. \textbf{83} (1951) 10.

%%%%%%%%%%%%%%%%%%%%%%%%%%%%%%%%%%%%%%%%%%
\bibitem{quantum-aspects-of-bhs_book}
X. Calmet,
``Quantum Aspects of Black Holes'',
Springer Int. Pub. (2015).

\bibitem{quantum-bh1}
X. Calmet and B. K. El-Menoufi,\\
``Quantum corrections to Schwarzschild black hole'',\\
Eur. Phys. J. C \textbf{77} (2017) 243.
[\href{https://arxiv.org/abs/1704.00261}{arXiv:1704.00261} [hep-th]].

\bibitem{quantum-bh2}
D. I. Kazakov and S. N. Solodukhin,\\
``On quantum deformation of the Schwarzschild solution'',\\
Nuc. Phys. B \textbf{429} (1994) 153.
[\href{https://arxiv.org/abs/hep-th/9310150}{arXiv:hep-th/9310150}].

\bibitem{quantum-bh3}
A. F. Ali and M. M. Khalil,
``Black hole with quantum potential'',\\
Nuc. Phys. B \textbf{909} (2016) 173.
[\href{https://arxiv.org/abs/1509.02495}{arXiv:1509.02495} [gr-qc]].

%%%%%%%%%%%%%%%%%%%%%%%%%%%%%%%%%%%%%%%%%%%%%%%%
\bibitem{bardeen-rbh}
J. M. Bardeen,
``Non-singular general-relativistic gravitational collapse'',
in Proceedings of International Conference GR5, Tbilisi, U.S.S.R. (1968).

\newpage
\bibitem{hayward-rbh}
S. A. Hayward,
``Formation and Evaporation of Nonsingular Black Holes'',\\
Phys. Rev. Lett. \textbf{96} (2006) 031103.
[\href{https://arxiv.org/abs/gr-qc/0506126}{arXiv:gr-qc/0506126}].

\bibitem{frolov-rbh}
V. P. Frolov,
``Information loss problem and a `black hole' model with a closed apparent horizon'',
J. High Energ. Phys. \textbf{2014} (2014).
[\href{https://arxiv.org/abs/1402.5446}{arXiv:1402.5446} [hep-th]].
%\enlargethispage{20pt}

\bibitem{rbhs-review}
S. Ansoldi,
``Spherical black holes with regular center: a review of existing models including a recent realization with Gaussian sources'',
(2008). \newline
[\href{https://arxiv.org/abs/0802.0330}{arXiv:0802.0330} [gr-qc]].

\bibitem{rbh-viability}
R. Carballo-Rubio, F. Di Filippo, S. Liberati, C. Pacilio and M. Visser,\\
``On the viability of regular black holes'',
J. High Energ. Phys. \textbf{2018} (2018).
[\href{https://arxiv.org/abs/1805.02675}{arXiv:1805.02675} [gr-qc]].

%%%%%%%%%%%%%%%%%%%%%%%%%%%%%%%%%%%%%%%%%%%%%%
\bibitem{morris-thorne-original}
M Morris and K. S. Thorne,
``Wormholes in spacetime and their use for interstellar travel: A tool for teaching General Relativity'',
Am. J. Phys. \textbf{56} (1988) 395.

\bibitem{morris-thorne-yurtsever}
M. S. Morris, K. S. Thorne and U. Yurtsever, 
``Wormholes, Time Machines, and the Weak Energy Condition?'',
Phys. Rev. Lett. \textbf{61} (1988) 1446.

\bibitem{lorentzian-wormholes}
M. Visser,
``Lorentzian Wormholes: From Einstein to Hawking'',\\
AIP press [now Springer], New York (1995).

\bibitem{visser-wormhole-examples}
M. Visser,
``Traversable wormholes: Some simple examples'',\\
Phys. Rev. D \textbf{39} (1989) 3182.
[\href{https://arxiv.org/abs/0809.0907}{arXiv:0809.0907} [gr-qc]].

\bibitem{visser-surgical}
M.~Visser,
``Traversable wormholes from surgically modified Schwarzschild space-times'',
Nucl. Phys. B \textbf{328} (1989), 203-212
doi:10.1016/0550-3213(89)90100-4
[\href{https://arxiv.org/abs/0809.0927}{arXiv:0809.0927} [gr-qc]].
%242 citations counted in INSPIRE as of 13 Jun 2020

\bibitem{baby}
M.~Visser,
``Wormholes, Baby Universes and Causality,''
Phys. Rev. D \textbf{41} (1990), 1116
doi:10.1103/PhysRevD.41.1116
%49 citations counted in INSPIRE as of 25 Aug 2020

\bibitem{visser-kar-dadhich}
M.~Visser, S.~Kar and N.~Dadhich,
``Traversable wormholes with arbitrarily small energy condition violations'',
Phys. Rev. Lett. \textbf{90} (2003), 201102
doi:10.1103/PhysRevLett.90.201102
[\href{https://arxiv.org/abs/gr-qc/0301003}{arXiv:gr-qc/0301003} [gr-qc]].
%239 citations counted in INSPIRE as of 13 Jun 2020

\bibitem{time-machine}
M.~Visser,
``From wormhole to time machine: Comments on Hawking's chronology protection conjecture,''
Phys. Rev. D \textbf{47} (1993), 554-565
doi:10.1103/PhysRevD.47.554
[\href{https://arxiv.org/abs/hep-th/9202090}{arXiv:hep-th/9202090} [hep-th]].
%47 citations counted in INSPIRE as of 25 Aug 2020

\bibitem{Kar:2004}
S.~Kar, N.~Dadhich and M.~Visser,
``Quantifying energy condition violations in traversable wormholes'',
Pramana \textbf{63} (2004), 859-864
doi:10.1007/BF02705207
[\href{https://arxiv.org/abs/gr-qc/0405103}{arXiv:gr-qc/0405103} [gr-qc]].
%57 citations counted in INSPIRE as of 21 Aug 2020

\bibitem{poisson-visser}
E.~Poisson and M.~Visser,
``Thin shell wormholes: Linearization stability'',\\
Phys. Rev. D \textbf{52} (1995), 7318-7321
doi:10.1103/PhysRevD.52.7318
[\href{https://arxiv.org/abs/gr-qc/9506083}{arXiv:gr-qc/9506083} [gr-qc]].
%236 citations counted in INSPIRE as of 13 Jun 2020

\bibitem{natural}
J.~G.~Cramer, R.~L.~Forward, M.~S.~Morris, M.~Visser, G.~Benford and G.~A.~Landis,
``Natural wormholes as gravitational lenses,''
Phys. Rev. D \textbf{51} (1995), 3117-3120
doi:10.1103/PhysRevD.51.3117
[\href{https://arxiv.org/abs/astro-ph/9409051}{arXiv:astro-ph/9409051} [astro-ph]].
%140 citations counted in INSPIRE as of 25 Aug 2020

\bibitem{R=0}
N.~Dadhich, S.~Kar, S.~Mukherji and M.~Visser,
``$R = 0$ space-times and selfdual Lorentzian wormholes,''
Phys. Rev. D \textbf{65} (2002), 064004
doi:10.1103/PhysRevD.65.064004
[\href{https://arxiv.org/abs/gr-qc/0109069}{arXiv:gr-qc/0109069} [gr-qc]].
%63 citations counted in INSPIRE as of 25 Aug 2020

\enlargethispage{20pt}
%=================================================
\bibitem{expmetric}
P. Boonserm, T. Ngampitipan, A. Simpson, and M. Visser,\\
``The exponential metric represents a traversable wormhole'',\\
Phys. Rev. D \textbf{98} (2018) 084048.
[\href{https://arxiv.org/abs/1805.03781}{arXiv:1805.03781} [gr-qc]].

\bibitem{blackbounce}
A. Simpson and M. Visser,
``Black-bounce to traversable wormhole'',\\
JCAP \textbf{1902} (2019) 042.
[\href{https://arxiv.org/abs/1812.07114}{arXiv:1812.07114} [gr-qc]].

\bibitem{blackbounce2}
A.~Simpson, P.~Mart\'in-Moruno and M.~Visser,
``Vaidya spacetimes, black-bounces, and traversable wormholes,''
Class. Quant. Grav. \textbf{36} (2019) no.14, 145007
doi:10.1088/1361-6382/ab28a5
[\href{https://arxiv.org/abs/1902.04232}{arXiv:1902.04232} [gr-qc]].
%12 citations counted in INSPIRE as of 25 Aug 2020

\bibitem{Lobo:2020}
F.~S.~N.~Lobo, A.~Simpson and M.~Visser,
``Dynamic thin-shell black-bounce traversable wormholes'',
Phys. Rev. D \textbf{101} (2020) no.12, 124035
doi:10.1103/PhysRevD.101.124035
[\href{https://arxiv.org/abs/2003.09419}{arXiv:2003.09419} [gr-qc]].
%6 citations counted in INSPIRE as of 20 Aug 2020


%\newpage
%%%%%%%%%%%%%%%%%%%%%%%%%%%%%%%%%%%%%%%%%%%
%%%%%%%%%%%%%%%%%%%%%%%%%%%%%%%%%%%%%%%%%%%
\bibitem{gravastar-original}
P. O. Mazur and E. Mottola,
``Gravitational vacuum condensate stars'',\\
Proceedings of the National Academy of Sciences \textbf{101} (2004) 9545.

\bibitem{gravastars-bh-alternative}
P. O. Mazur and E. Mottola,\\
``Gravitational Condensate Stars: An Alternative to Black Holes'', (2001).\\{}
[\href{https://arxiv.org/abs/gr-qc/0109035}{arXiv:0109035} [gr-qc]]

\bibitem{visser-gravastars}
M. Visser and D. Wiltshire,
``Stable gravastars: An alternative to black holes?'',
Class. Quant. Grav. \textbf{21} (2004) 1135.
[\href{https://arxiv.org/abs/gr-qc/0310107}{arXiv:gr-qc/0310107}].

\bibitem{anisotropic}
C.~Catto\"en, T.~Faber and M.~Visser,
``Gravastars must have anisotropic pressures,''
Class. Quant. Grav. \textbf{22} (2005), 4189-4202
doi:10.1088/0264-9381/22/20/002
[arXiv:gr-qc/0505137 [gr-qc]].
%165 citations counted in INSPIRE as of 25 Aug 2020

\bibitem{lobo-gravastars}
F. S. N. Lobo,
``Stable dark energy stars'',
Class. Quant. Grav. \textbf{23} (2006) 1525.
[\href{https://arxiv.org/abs/gr-qc/0508115}{arXiv:gr-qc/0508115}].

\bibitem{MartinMoruno:2011}
P.~Mart\'in-Moruno, N.~Montelongo-Garc\'ia, F.~S.~N.~Lobo and M.~Visser,
``Generic thin-shell gravastars,''
JCAP \textbf{03} (2012), 034
doi:10.1088/1475-7516/2012/03/034
[\href{https://arxiv.org/abs/1112.5253}{arXiv:1112.5253} [gr-qc]].
%31 citations counted in INSPIRE as of 25 Aug 2020

\bibitem{Lobo:2015}
F.~S.~N.~Lobo, P.~Mart\'in-Moruno, N.~Montelongo-Garc\'ia and M.~Visser,
``Novel stability approach of thin-shell gravastars,''
doi:10.1142/9789813226609\_0221
[\href{https://arxiv.org/abs/1512.07659}{arXiv:1512.07659} [gr-qc]].
%8 citations counted in INSPIRE as of 25 Aug 2020

%%%%%%%%%%%%%%%%%%%%%%%%%%%%%%%%%%%%%%%%%%%%%

\bibitem{Cunha1}
P.~Cunha, V.P., E.~Berti and C.~A.~R.~Herdeiro,\\
``Light-Ring Stability for Ultracompact Objects'',\\
Phys. Rev. Lett. \textbf{119} (2017) no.25, 251102
doi:10.1103/PhysRevLett.119.251102
[\href{https://arxiv.org/abs/1708.04211}{arXiv:1708.04211} [gr-qc]].
%59 citations counted in INSPIRE as of 07 Sep 2020

\bibitem{Cunha2}
P.~Cunha, V.P. and C.~A.~R.~Herdeiro,\\
``Stationary black holes and light rings'',\\
Phys. Rev. Lett. \textbf{124} (2020) no.18, 181101
doi:10.1103/PhysRevLett.124.181101
[\href{https://arxiv.org/abs/2003.06445}{arXiv:2003.06445} [gr-qc]].
%2 citations counted in INSPIRE as of 07 Sep 2020

%%%%%%%%%%%%%%%%%%%%%%%%%%%%%%%%%%%%%%%%%%%%%

\bibitem{pandora}
R.~Carballo-Rubio, F.~Di Filippo, S.~Liberati and M.~Visser,
``Opening the Pandora's box at the core of black holes,''
Class. Quant. Grav. \textbf{37} (2020) no.14, 145005
doi:10.1088/1361-6382/ab8141
[\href{https://arxiv.org/abs/1908.03261}{arXiv:1908.03261} [gr-qc]].
%17 citations counted in INSPIRE as of 25 Aug 2020

\bibitem{small-dark-heavy}
M.~Visser, C.~Barcel\'o, S.~Liberati and S.~Sonego,\\
``Small, dark, and heavy: But is it a black hole?,''\\
PoS \textbf{BHGRS} (2008), 010
doi:10.22323/1.075.0010
[\href{https://arxiv.org/abs/0902.0346}{arXiv:0902.0346} [gr-qc]].
%44 citations counted in INSPIRE as of 25 Aug 2020

\bibitem{observability}
M.~Visser,
``Physical observability of horizons,''
Phys. Rev. D \textbf{90} (2014) no.12, 127502
doi:10.1103/PhysRevD.90.127502
[\href{https://arxiv.org/abs/1407.7295}{arXiv:1407.7295} [gr-qc]].
%40 citations counted in INSPIRE as of 25 Aug 2020

\bibitem{visser-phenom-aspects}
R. Carballo-Rubio, F. Di Filippo, S. Liberati and M. Visser,
``Phenomenological aspects of black holes beyond general relativity'',
Phys. Rev. D \textbf{98} (2018) 124009.
[\href{https://arxiv.org/abs/1809.08238}{arXiv:1809.08238} [gr-qc]].

%\newpage
%%%%%%%%%%%%%%%%%%%%%%%%%%%%%%%%%%%%%%%%%%%%
\bibitem{eht-1}
The Event Horizon Telescope Collaboration,
``First M87 Event Horizon Telescope Results. I. The Shadow of the Supermassive Black Hole'',
ApJL \textbf{875} (2019) L1.
[\href{https://arxiv.org/abs/1906.11238}{arXiv:1906.11238} [astro-ph.GA]].

\bibitem{eht-2}
The Event Horizon Telescope Collaboration,
``First M87 Event Horizon Telescope Results. II. Array and Instrumentation'',
ApJL \textbf{875} (2019) L2. \newline
[\href{https://arxiv.org/abs/1906.11239}{arXiv:1906.11239} [astro-ph.GA]].

\bibitem{eht-3}
The Event Horizon Telescope Collaboration,
``First M87 Event Horizon Telescope Results. III. Data Processing and Calibration'',
ApJL \textbf{875} (2019) L3.
[\href{https://arxiv.org/abs/1906.11240}{arXiv:1906.11240} [astro-ph.GA]].

\bibitem{eht-4}
The Event Horizon Telescope Collaboration,
``First M87 Event Horizon Telescope Results. IV. Imaging the Central Supermassive Black Hole'',
ApJL \textbf{875} (2019) L4.
[\href{https://arxiv.org/abs/1906.11241}{arXiv:1906.11241} [astro-ph.GA]].

\bibitem{eht-5}
The Event Horizon Telescope Collaboration,
``First M87 Event Horizon Telescope Results. V. Physical Origin of the Asymmetric Ring'',
ApJL \textbf{875} (2019) L5.
[\href{https://arxiv.org/abs/1906.11242}{arXiv:1906.11242} [astro-ph.GA]].

\bibitem{eht-6}
The Event Horizon Telescope Collaboration,
``First M87 Event Horizon Telescope Results. VI. The Shadow and Mass of the Central Black Hole'',
ApJL \textbf{875} (2019) L6.
[\href{https://arxiv.org/abs/1906.11243}{arXiv:1906.11243} [astro-ph.GA]].

%%%%%%%%%%%%%%%%%%%%%%%%%%%%%%%%%%%%%%%%%%%%%
\bibitem{ligo-detection-papers}
See \url{https://www.ligo.caltech.edu/page/detection-companion-papers} for a collection of detection papers from LIGO. \\
See also \url{https://pnp.ligo.org/ppcomm/Papers.html} for a complete list of publications from the LIGO Scientific Collaboration and Virgo Collaboration.

\bibitem{grav-wave-observations-wiki}
See, for example, \href{https://en.wikipedia.org/wiki/List_of_gravitational_wave_observations}{wikipedia.org/List\_of\_gravitational\_wave\_observations} for a list of current (May 2020) gravitational wave observations.
\enlargethispage{20pt}


%%%%%%%%%%%%%%%%%%%%%%%%%%%%%%%%%%%%%%%
\bibitem{LISA}
E.~Barausse, E.~Berti, T.~Hertog, S.~A.~Hughes, P.~Jetzer, P.~Pani, T.~P.~Sotiriou, N.~Tamanini, H.~Witek, K.~Yagi, N.~Yunes, \emph{et al.},\\
``Prospects for Fundamental Physics with LISA'',\\
Gen.Rel.Grav. \textbf{52} (2020) 8, 
doi:10.1007/s10714-020-02691-1.\\{}
[\href{https://arxiv.org/abs/2001.09793}{arXiv:2001.09793} [gr-qc]].
%16 citations counted in INSPIRE as of 10 Jun 2020

\newpage
%%%%%%%%%%%%%%%%%%%%%%%%%%%%%%%%%%%%%%%%%%%%%%
\bibitem{asymptot-mink-core}
A. Simpson and M. Visser,\\\
``Regular black holes with asymptotically Minkowski cores'',\\
Universe \textbf{6} (2020) 8.
[\href{https://arxiv.org/abs/1911.01020}{arXiv:1911.01020} [gr-qc]].

\bibitem{Berry:2020}
T.~Berry, F.~S.~N.~Lobo, A.~Simpson and M.~Visser,
``Thin-shell traversable wormhole crafted from a regular black hole with asymptotically Minkowski core'', Physical Review D (in press), 
[\href{https://arxiv.org/abs/2008.07046}{arXiv:2008.07046} [gr-qc]].
%0 citations counted in INSPIRE as of 20 Aug 2020

%%%%%%%%%%%%%%%%%%%%%%%%%%%%%%%%%%%%%%%%%%%
%%%%%%%%%%%%%%%%%%%%%%%%%%%%%%%%%%%%%%%%%%%%
\bibitem{Culetu:2013}
  H.~Culetu,
  ``On a regular modified Schwarzschild spacetime'',\\
  \href{https://arxiv.org/abs/1305.5964}{arXiv:1305.5964} [gr-qc].
  %%CITATION = ARXIV:1305.5964;%%
  %11 citations counted in INSPIRE as of 17 Nov 2019
  
  %\newpage
  \bibitem{Culetu:2014}
  H.~Culetu,
  ``On a regular charged black hole with a nonlinear electric source'',\\
  Int.\ J.\ Theor.\ Phys.\  {\bf 54} (2015) no.8,  2855
  doi:10.1007/s10773-015-2521-6
  [\href{https://arxiv.org/abs/1408.3334}{arXiv:1408.3334} [gr-qc]].
  %%CITATION = doi:10.1007/s10773-015-2521-6;%%
  %22 citations counted in INSPIRE as of 17 Nov 2019

%\newpage
 \bibitem{Culetu:2015a}
  H.~Culetu,
  ``Nonsingular black hole with a nonlinear electric source'',\\
  Int.\ J.\ Mod.\ Phys.\ D {\bf 24} (2015) no.09,  1542001.
  doi:10.1142/S0218271815420018
  %%CITATION = doi:10.1142/S0218271815420018;%%
  %4 citations counted in INSPIRE as of 17 Nov 2019 
  
  \bibitem{Culetu:2015b}
  H.~Culetu,
  ``Screening an extremal black hole with a thin shell of exotic matter'',\\
  Phys.\ Dark Univ.\  {\bf 14} (2016) 1
  doi:10.1016/j.dark.2016.07.004\\{}
  [\href{https://arxiv.org/abs/1508.01102}{arXiv:1508.01102} [gr-qc]].
  %%CITATION = doi:10.1016/j.dark.2016.07.004;%%
  %4 citations counted in INSPIRE as of 17 Nov 2019
  
  \bibitem{Junior:2015}
  E.~L.~B.~Junior, M.~E.~Rodrigues and M.~J.~S.~Houndjo,\\
  ``Regular black holes in $f(T)$ Gravity through a nonlinear electrodynamics source'',
  JCAP {\bf 1510} (2015) 060
  doi:10.1088/1475-7516/2015/10/060\\{}
  [\href{https://arxiv.org/abs/1503.07857}{arXiv:1503.07857} [gr-qc]].
  %%CITATION = doi:10.1088/1475-7516/2015/10/060;%%
  %46 citations counted in INSPIRE as of 18 Nov 2019
  
  \bibitem{Rodrigues:2015}
  M.~E.~Rodrigues, E.~L.~B.~Junior, G.~T.~Marques and V.~T.~Zanchin,\\
  ``Regular black holes in $f(R)$ gravity coupled to nonlinear electrodynamics'',\\
  Phys.\ Rev.\ D {\bf 94} (2016) no.2,  024062\\
   Addendum: [Phys.\ Rev.\ D {\bf 94} (2016) no.4,  049904]
  doi:10.1103/PhysRevD.94.024062, 10.1103/PhysRevD.94.049904\\{}
  [\href{https://arxiv.org/abs/1511.00569}{arXiv:1511.00569} [gr-qc]].
  %%CITATION = doi:10.1103/PhysRevD.94.024062, 10.1103/PhysRevD.94.049904;%%
  %25 citations counted in INSPIRE as of 18 Nov 2019
  
%==================================================================

\bibitem{Corless:1996}
R.~M.~Corless, G.~H.~Gonnet, D.~E.~G.~Hare, D.~J.~Jeffrey and D.~E.~Knuth,\\
``On the Lambert $W$ function'',
Adv. Comput. Math. \textbf{5} (1996), 329-359
doi:10.1007/BF02124750
%286 citations counted in INSPIRE as of 22 Aug 2020

\bibitem{OSCO}
P.~Boonserm, T.~Ngampitipan, A.~Simpson and M.~Visser,
``Innermost and outermost stable circular orbits in the presence of a positive cosmological constant'',
Phys. Rev. D \textbf{101} (2020) no.2, 024050
doi:10.1103/PhysRevD.101.024050
[\href{https://arxiv.org/abs/1909.06755}{arXiv:1909.06755} [gr-qc]].
%2 citations counted in INSPIRE as of 25 Aug 2020

%==================================================================
\end{thebibliography}
\end{document}